\documentclass[12pt,preprint]{aastex}

\newcommand{\msun}{\mbox{ M$_{\odot}$}}
\newcommand{\lbsun}{\mbox{ L$_{B\odot}$}}
\newcommand{\bq}{\begin{equation}}
\newcommand{\eq}{\end{equation}}
\newcommand{\kpc}{\mbox{ kpc}}
\newcommand{\Mpc}{\mbox{ Mpc}}
\newcommand{\hunits}{\mbox{ km s$^{-1}$ Mpc$^{-1}$}}
\newcommand{\yr}{\mbox{ yr}}
\newcommand{\kel}{\mbox{ K}}
\newcommand{\cmden}{\mbox{ cm$^{-3}$}}
\newcommand{\secinv}{\mbox{ s$^{-1}$}}
\newcommand{\cmdensec}{\mbox{ cm$^{-3}$  s$^{-1}$}}
\newcommand{\bkgdflux}{\mbox{ photons cm$^{-2}$  s$^{-1}$ keV$^{-1}$
sr$^{-1}$}} 
\newcommand{\photflux}{\mbox{ photons cm$^{-2}$  s$^{-1}$}} 
\newcommand{\ergs}{\mbox{ erg s$^{-1}$}}

\newcommand{\keV}{\mbox{ keV}}

\newcommand{\microgauss}{\mbox{ $\mu$G}}
\newcommand{\mubarn}{\mbox{ $\mu$barn}}
\newcommand{\mbarn}{\mbox{ mbarn}}
\newcommand{\mach}{{\mathcal M}}

\begin{document}
 
\title{Emission of Positron Annihilation Line Radiation \\ by
Clusters of Galaxies} 

\author{Steven R. Furlanetto \& Abraham Loeb}
\affil{Harvard-Smithsonian Center for Astrophysics, 60 Garden St.,
Cambridge, MA 02138;\\
sfurlanetto@cfa.harvard.edu, aloeb@cfa.harvard.edu}

\begin{abstract}

Clusters of galaxies are enriched with positrons from jets of active
galactic nuclei (AGNs) or from the interaction of cosmic-rays with the
intracluster gas.  We follow the cooling of these positrons and show that
their eventual annihilation with cluster electrons yields a narrow
annihilation line.  Unlike annihilation in the interstellar medium of
galaxies, the line produced in clusters is not smeared by three-photon
decay of positronium, because positronium formation is suppressed at
the high ($\ga 1$ keV) temperature of the cluster electrons. We show
that if AGN jets are composed of $e^+e^-$ pairs, then the
annihilation line from rich clusters within a distance of 100 Mpc
might be detectable with future space missions, such as INTEGRAL or
EXIST. 

\end{abstract}


\section{ Introduction }

Despite decades of intense study, the composition of relativistic radio
jets remains enigmatic.  While the existence of synchrotron-emitting
electrons is secure in both quasars \citep{bbr} and microquasars
\citep{mirabel}, there is no conclusive evidence that can determine whether
positrons or protons make up the positively charged component.  Naively,
one might expect these cases to be easily distinguishable through an
observational search for the electron-positron annihilation line.  However,
modeling of the annihilation process in active jets has shown this not to
be the case: the annihilation spectral feature is not a line but is instead
very broad and hence difficult to unambiguously identify (although it may
possibly contribute to the observed $\gamma$-ray spectrum of blazars; see
\citealt{boettcher}).  Even if positron cooling to non-relativistic
energies within the jet is efficient, the high bulk Lorentz factor
($\gamma_{\rm jet} \sim 10$) of the jet would likely smear out an
annihilation line \citep{bbr}.  The resulting radiation may be detectable
with upcoming space observatories, although its interpretation will depend
critically on proper modeling of the region inside the jet \citep{wang}.

The chief obstacle to producing a recognizable annihilation-line feature is
the highly relativistic nature of the associated plasma.  This obstacle may
be overcome simply by waiting for the active galactic nucleus (AGN) to
become dormant.  Once the central engine disappears, the material in the
jet will presumably mix with the ambient medium and cool to the ambient
temperature.  In this paper, we calculate the annihilation signal from
``relic'' positrons produced by an AGN embedded in a galaxy cluster, as the
positrons thermalize with the electrons in the intracluster medium (ICM).
We will show that under reasonable assumptions an annihilation line from
nearby clusters would be observable in the near future.

A similar suggestion of thermal annihilation after escape from the
accelerating source was made by \citet{maciolek} in the context of
small-scale jets from stellar-mass black holes.  However, in galactic
environments with typical temperatures $T \la 10^6 \kel$, formation of an
annihilation line at $511 \keV$ is inhibited by the rapid formation of
positronium, whose primary annihilation channel yields three photons.  For
example, the observations of \citet{kinzer} show that $93 \pm 4 \%$ of pair
annihilations in the Galactic center occur through the positronium channel
and that $\ga 70\%$ of the total annihilation energy is emitted in a broad
continuum rather than in the line.  We will argue that galaxy clusters are
ideally suited to producing annihilation lines because the characteristic
temperature of the ICM is larger than the binding energy of positronium;
hence, nearly all annihilations produce photons near $511 \keV$.

Another important open question involves the dynamical significance of
cosmic rays to the ICM.  Faraday rotation studies have revealed that
cluster cores have magnetic fields $\ga 1 \microgauss$ \citep{kim,clarke},
while observations of excess emission in the ultraviolet and hard X-ray
bands from the ICM have been used to infer the existence of widespread
cluster fields at slightly lower levels $\sim 0.1$--$1 \microgauss$ (e.g.,
\citealt{rephaeli2}).  The pervasiveness of magnetic fields in clusters
indicates that collisionless shocks generated by accretion or merger events
may be efficient particle accelerators \citep{bland-eichler,col}.
Observations of synchrotron radio halos indicate that the acceleration of
electrons by cluster merger shocks is a common occurrence \citep{kemp-sar}.
However, these observations probe only the electron component.  In order to
understand the dynamical effects of the cosmic rays on the cluster, we are
most interested in the accelerated protons, both because shock acceleration
may inject more energy into this component than into the electron
component \citep{fields,but} and because cosmic ray protons do not
rapidly lose their energy through radiative cooling.

Recent cosmological simulations suggest that the cosmic ray pressure might
be as large as $\sim 10$--$40 \%$ of the thermal gas pressure in clusters
\citep{miniati-prot}.  To date, two diagnostics of the cosmic ray proton
content have been proposed, both relying on the decay of pions produced in
collisions between cosmic ray and thermal protons, namely synchrotron and
inverse Compton emission from secondary electrons and positrons produced in
charged pion decay \citep{blasi,dolag,miniati-sec} and $\gamma$-rays
produced in neutral pion decay \citep{col,miniati-prot}.  The former
process suffers from the possibility of contamination by newly
accelerated electrons, while the $\gamma$-ray signal will remain below the
detection limits at least until the launch of the GLAST
mission\footnote{See http://www-glast.stanford.edu/mission.html} in 2005,
largely because the energy from the decaying pions is distributed over a
very wide range of photon energies.

Because of the rapid cooling of positrons produced in the decay of $\pi^+$
to non-relativistic temperatures, we would expect an annihilation line to
be produced in this case as well.  (Here too the formation of positronium
is inhibited by the relatively large cluster temperatures.)  We therefore
also calculate the annihilation spectrum produced through secondary
positron production by cosmic ray protons.  However, we find that this
signal is well below the detection thresholds of upcoming instruments.
Hence, AGN are the only realistic pollutant of substantial amounts of
positrons into galaxy clusters.  We thus argue that the future detection of
positron annihilation line radiation from clusters would constitute a
robust signature of electron-positron jets.

We begin by describing the factors determining the evolution of the
positron population in \S \ref{posdfevol}, including cooling, annihilation,
and source terms.  We then solve the evolution equations in \S
\ref{posdfsoln}.  In \S \ref{annspec} we calculate the resulting positron
annihilation signals.  Finally, we conclude in \S \ref{conc} with a
discussion of our results and prospects for future observations.

Throughout the paper, we assume a $\Lambda$CDM cosmology with $\Omega_0 =
0.3$, $\Omega_\Lambda = 0.7$, and $H_0 = 70 \hunits$.  

\section{ Evolution of the Positron Distribution Function }
\label{posdfevol}

The time evolution of $N_+(\gamma,t)$, the mean
differential number density of positrons with Lorentz factor between
$(\gamma, \gamma + d\gamma)$ in the cluster at time $t$, is described by the
analog of the 
Boltzmann equation with source and sink terms (\citealt{sarazin}): 
\bq
\frac{\partial N_+(\gamma,t)}{\partial t} = \frac{\partial}{\partial
\gamma}[b(\gamma,t)\, N_+(\gamma,t)] + Q(\gamma) - N_+(\gamma,t) A(\gamma).
\label{eq:dfevol}
\eq 
Here, $b(\gamma,t)=(d\gamma/dt)$~ is the cooling rate of a single
positron, $Q(\gamma)d\gamma$ is the production rate of positrons per unit
volume in the interval $(\gamma,\gamma+d\gamma)$, and $A(\gamma)d\gamma$ is
the annihilation rate of positrons in the same interval.  We have neglected
diffusion and loss of positrons between volume elements in the ICM
\citep{ginzburg}.  This should be a good approximation provided that the
sources of positrons are distributed over a sufficiently large volume, as
would be expected for injection by both AGNs and cluster shocks.
Alternatively, because clusters trap all but the very highest energy cosmic
rays over a Hubble time \citep{berezinsky}, our evolution equation can be
viewed as simply describing the total positron population in the cluster,
with the appropriate averages of the cooling, source, and annihilation
terms \citep{sarazin}.

We are concerned primarily with those positrons that cool sufficiently
to contribute to line emission upon annihilation.  Therefore, we wish
to track the number of positrons that have thermalized with the
ambient electrons.  In the test particle approximation, a positron
cools until its kinetic energy $\epsilon_{K}
\approx 0.98 k_B T_e$, where $k_B$ is Boltzmann's constant and $T_e$ is
the temperature of the ambient thermal electrons \citep{trubnikov}.
In actuality, the finite population of cooled positrons thermalizes
with the ambient electrons on a timescale \citep{trubnikov} 
\bq 
\tau_{\rm therm} \approx 4.8 \times 10^3 \, T_{\rm keV}^{3/2} \left(
\frac{10^{-3} \cmden}{n_e} \right) \yr,
\label{eq:ttherm}
\eq 
where $n_e$ is the electron number density in the cluster core and
$T_{\rm keV} = (k_B T_e/{\rm keV})$.  Because $\tau_{\rm therm}$ is much
shorter than the other relevant timescales, we assume that thermalization
occurs instantaneously.  The pool of cooled positrons essentially acts as
an absorbing wall in the cooling equation, so we write the total positron
distribution function $N_T(\gamma)$ as \bq N_T(\gamma,t) = n_+(t)
\delta(\gamma-\gamma_{\rm eq}) + N_+(\gamma,t),
\label{eq:dftot}
\eq 
where $\gamma_{\rm eq}$ is the mean Lorentz factor of the ambient
electrons and $n_+$ is the number density of the thermalized
positrons.

In the following subsections, we describe each of the terms on the
right-hand side of equation (\ref{eq:dfevol}) in detail.  We solve for
the $N_+(\gamma,t)$ appropriate to our two models in \S \ref{posdfsoln}.  

\subsection{ The Loss Rate $b(\gamma,t)$ }

The principal cooling mechanisms for cosmic ray electrons or
positrons in a cluster are synchrotron, inverse Compton (IC), and Coulomb
cooling \citep{rephaeli}.  The synchrotron and IC loss rates for a single
positron with Lorentz factor $\gamma\gg 1$ are given by \citep{rephaeli}
\bq 
\left( \frac{d \gamma}{d t} \right)_{\rm syn,IC} = - \frac{4
\sigma_T}{3 m_e c} (u_B + u_{\rm CMB}) \gamma^2,
\label{eq:bsynic1}
\eq
where $\sigma_T$ is the Thomson cross section, $m_e$ is the electron
mass, $u_B = B^2/8 \pi$ is the energy density in the cluster magnetic
field, and $u_{\rm CMB}$ is the energy density of the cosmic microwave
background (CMB).  The loss rate due to these two processes is therefore
\bq
b_{\rm syn,IC}(\gamma,t) = - \left( \frac{d \gamma}{d t}  \right)_{\rm
syn,IC} = 4.3 \times 10^{-13} \, \gamma^2 
\left[ (1+z)^4 + 0.86 \left( \frac{B}{3 \microgauss} \right)^2 \right]
\yr^{-1},
\label{eq:bsynic2}
\eq
where $z$ is the cosmological redshift at cosmic time $t$.

The loss rate from Coulomb cooling is approximately \citep{kempner}
\bq
\left( \frac{d \gamma}{d t} \right)_{\rm Coul} = - \frac{4 \pi n_e
e^4}{\beta m_e^2 c^3} \ln \left( \frac{1.12 m_e c^2 \gamma^{1/2}
\beta^2}{\hbar \omega_p} \right),
\label{eq:bcoul1}
\eq
where $\beta = (1 - \gamma^{-2})^{1/2}$ is the positron speed
normalized to the speed of light and $\omega_p$ is the plasma
frequency of the ambient electron gas.  The corresponding
choice for the Coulomb
logarithm is valid for $\alpha \ll \beta \ll 1$, where $\alpha$ is the
fine structure constant, and for $\gamma \gg 1$, with errors $\la
10\%$ in the intermediate regime \citep{kempner}.  Evaluating the
constants, we find
\bq
b_{\rm Coul}(\gamma) = - \left( \frac{d \gamma}{d t}  \right)_{\rm
Coul} = 3.5 \times 10^{-5} \frac{n_{\rm cm}}{\beta} \left[ 1 +
\frac{1}{74.5} \ln \left( \frac{\gamma \beta^4}{n_{\rm cm}} \right)
\right] \yr^{-1}, 
\label{eq:bcoul2}
\eq
where $n_{\rm cm} = (n_e/{\rm cm}^{-3})$.

The total loss rate is then $b(\gamma,t) = b_{\rm Coul}(\gamma) + b_{\rm
syn,IC}(\gamma,t)$.  Comparing equations (\ref{eq:bsynic2}) and
(\ref{eq:bcoul2}), we see that IC/synchrotron cooling dominates for
$\gamma \ga 200 (n_e/10^{-3} \cmden)^{1/2}$; hence our assumption that
$\gamma \gg 1$ in equation (\ref{eq:bsynic1}) is appropriate.

\subsection{ The Annihilation Rate $A(\gamma)$ }

The annihilation rate $A(\gamma_+)$ of positrons at a given Lorentz
factor through the process $e^+e^- \rightarrow 2\gamma$ may be written
as \citep{svensson} 
\bq
A(\gamma_+) = \int_1^\infty d\gamma_- N_- (\gamma_-) \int_{-1}^{1}
\frac{d\mu}{2} \frac{\gamma_{cm}^2}{\gamma_+ \gamma_-} 2 \beta_{cm} c
\sigma_{ee}(\gamma_{cm}), 
\label{eq:posanndefn}
\eq 
assuming that the distribution function of electrons, $N_-(\gamma_-)$,
is isotropic but otherwise arbitrary.  Here $\gamma_+$ $(\gamma_-)$ is the
Lorentz factor of the positron (electron) in the cluster rest-frame,
$\gamma_{cm}$ is the Lorentz factor of the positron in the center-of-mass
(CM) frame, $\mu$ is the interaction angle in the cluster rest-frame,
and $\sigma_{ee}(\gamma_{cm})$ 
is the total annihilation cross section in the CM frame.  We find that
assuming that the ambient electrons are all at rest [$N_-(\gamma_-) = n_e
\delta(\gamma_- - 1)$] makes only a small difference to the total
annihilation cross-section for the non-thermalized positrons, because
realistic cluster temperatures are highly non-relativistic (so that
$\gamma_+ \gg \gamma_-$).  In this ``{\it cosmic ray approximation},''
$\gamma_{cm}^2 = (\gamma_++1)/2$ and the annihilation rate becomes 
\bq 
A(\gamma_+) = \frac{\pi c
n_e r_e^2 \beta_+}{\gamma_+ + 1} \left\{ \frac{\gamma_+^2 + 4 \gamma_+ +
1}{\gamma_+^2 - 1} \ln \left[\gamma_+ + (\gamma_+^2 - 1)^{1/2} \right] -
\frac{\gamma_+ + 3}{(\gamma_+^2 - 1)^{1/2}} \right\},
\label{eq:posanncr}
\eq
where $r_e$ is the classical electron radius.  This expression is valid
provided Coulomb corrections may be neglected, i.e. as long as $\beta_+ \gg
\alpha$; this condition is satisfied for all $\gamma_+ \geq \gamma_{\rm
eq}$ at the typical cluster temperatures of $k_B T_e \ga 1 \keV$.

In order to calculate the spectrum of annihilation photons, we are also
interested in the derivative of the annihilation rate with respect to
photon energy.  This may be written in the {\it cosmic ray 
approximation} as
\citep{svensson}
\bq \frac{d n_\gamma}{dk\,dt} = n_e \int_1^\infty
d\gamma_+ N(\gamma_+) \langle v \frac{d \sigma}{d k} (k,\gamma_+)
\rangle,
\label{eq:posannspec}
\eq
where $k \equiv \epsilon_\gamma/m_e c^2$ is the normalized photon
energy and
\bq
\langle v \frac{d \sigma}{d k} (k,\gamma_+) \rangle = \frac{\pi c
r_e^2}{\beta_+ \gamma_+^2} 
\left\{ \frac{-(3+\gamma_+)/(1+\gamma_+) +
(3+\gamma_+)/k-1/k^2}{[1-k/(1+\gamma_+)]^2} -2 \right\}
\label{eq:vdsdk}
\eq
is the angle-averaged emissivity per electron-positron pair.

For the thermalized positrons, the {\it cosmic ray approximation} is 
invalid
because $\gamma_+ \approx \gamma_-$.  We therefore must evaluate the
general expression for the photon production rate,
\bq
\frac{d n_\gamma}{dk\,dt} = \int_1^\infty d\gamma_+ N(\gamma_+)
\int_1^\infty d\gamma_- N_-(\gamma_-) 
\langle v \frac{d \sigma}{d k} (k,\gamma_+,\gamma_-) \rangle,
\label{eq:posannemiss}
\eq
for Maxwell-Boltzmann electron and positron distributions at a
temperature $T_e$.  \citet{sven96} show that the result is
\bq
\left. \frac{d n_\gamma}{dk\,dt} \right|_{\rm line} = 
\frac{3}{4\sqrt{\pi}} n_+ n_e c \sigma_T
\, k^{3/2} \left( \frac{m_e c^2} { k_B T_e} \right)^{1/2} \exp \left[ -
\frac{(k-1)^2}{(k_B T_e/m_e c^2) \, k} \right]
\label{eq:posanntherm}
\eq
in the approximation $k \ll 1$ and $k_B T_e \ll m_e c^2$.  Here we use
the subscript ``line'' because annihilation of the thermalized component
yields a well-defined spectral line (see \S 4 below).  The total
annihilation rate of the thermalized positrons is then simply 
\begin{eqnarray}
A(\gamma_{\rm eq}) & = & \frac{1}{2 n_+} \int dk 
\left. \frac{d n_\gamma}{dk\,dt} \right|_{\rm line} \\
\,& \approx & 8 \times 10^{-15} n_e \, \secinv, \nonumber
\label{eq:ageq}
\end{eqnarray}
with only a very weak dependence on temperature in the range of
interest.  (The factor of two in the denominator appears because each
annihilating positron yields two photons.)

Note that in many astrophysical situations, positron annihilation occurs
principally through the formation and subsequent decay of positronium.  In
such situations, the dominant annihilation mechanism is the decay of
orthopositronium, which must emit at least three photons in order to
conserve angular momentum \citep{landau}.  This mechanism therefore
precludes the formation of an annihilation line.  Fortunately, in our case
positronium formation is insignificant.  For temperatures $k_B T_e \ga 1
\keV$, the cross section for positronium formation amounts to $\la 10\%$
of the cross section for direct annihilation, and the positronium formation
rate decreases strongly at yet higher temperatures
\citep{crannell}.  We therefore neglect this annihilation channel in our
calculations.  Note, however, that in cool galactic environments ($T \la
10^5 \kel$), annihilation through positronium formation dominates and the
annihilation line is suppressed.

\subsection{ The Source Function $Q(\gamma)$ }

In the case of positron injection purely via an AGN outflow, the source
function $Q(\gamma)$ is nonzero only at the instant of injection.  In this
section we describe positron production mechanisms in galaxy clusters
over and above direct injection from AGN.  The dominant mechanism, the
proton-proton channel, is described in \S \ref{ppcoll}.  For
completeness, we examine other mechanisms in \S \ref{posprodother}.

\subsubsection{ Positron Production from Proton-Proton Collisions }
\label{ppcoll}

Collisions between thermal and cosmic ray protons produce
positrons primarily through the reaction chain
\begin{eqnarray}
p + p & \rightarrow & \pi^+ + X, \nonumber \\
\pi^+ & \rightarrow & \mu^+ + \nu_\mu, \nonumber \\
\mu^+ & \rightarrow & e^+ + \bar{\nu}_\mu + \nu_e, \nonumber
\label{eq:ppchain}
\end{eqnarray}
provided that the cosmic ray proton has a Lorentz
factor exceeding $\gamma_{\rm th} = 1.3$, the threshold for pion
production \citep{mannheim}. 
The resulting positron production rate is then \citep{mosk}
\bq
Q_{pp}(\gamma) = n_H \int_{\gamma_p^{\rm min}}^\infty
d\gamma_p J_p(\gamma_p) \sigma_{\pi^+}(\gamma_p)
\int_{\gamma_\pi^{\rm min}}^{\gamma_{\pi}^{\rm max}} d
\gamma_{\pi} F_+(\gamma, \gamma_{\pi})
F_\pi(\gamma_{\pi}, \gamma_p),
\label{eq:posproddf}
\eq
where $n_H = n_e$ is the proton density in the cluster core,
$J_p(\gamma_p)$ is the differential cosmic ray proton flux at
Lorentz factor $\gamma_p$, $\sigma_{\pi^+}(\gamma_p)$ is the
cross-section for $\pi^+$ production in the collision (including all
channels), $F_+(\gamma, \gamma_{\pi})$ is the normalized
positron distribution function at positron Lorentz factor $\gamma$
for positrons produced by the decay of a pion with Lorentz factor
$\gamma_\pi$, and $F_\pi(\gamma_{\pi},\gamma_p)$ is the normalized
pion distribution function produced by a collision between a thermal
proton and a cosmic ray proton with Lorentz factor $\gamma_p$.  The
maximum cosmic ray energy $\gamma_p^{\rm max}$ is determined below.
The other integration limits are determined by kinematics:
$\gamma_\pi^{\rm min}$ is the minimum pion Lorentz factor needed to
produce a positron with $\gamma$, $\gamma_\pi^{\rm max}$ is the
maximum pion Lorentz factor that can be produced by a proton with
$\gamma_p$, and $\gamma_p^{\rm min}$ is the minimum proton Lorentz
factor required to produce a positron with $\gamma$.  

Our calculation of the resulting distribution function follows closely
the method outlined in Appendices B and C of \citet{mosk}, so we only
summarize it here.  Unfortunately, there is no firm theoretical
understanding of pion production in proton-proton collisions.  The
production cross-sections are experimentally well-determined (see the
compilation of \citealt{dermer}), but the resulting
$F_\pi(\gamma_\pi,\gamma_p)$ are not.  Instead, we must use
approximate models of the interaction.  At low energies ($\gamma_p
\la 3$), the ``isobaric model'' of \citet{stecker}, in which 
the collision forms a $\Delta$ isobar which subsequently decays
into a pion, fits the data best, while at high energies ($\gamma_p
\ga 7$), the ``scaling model'' of \citet{badhwar} provides a
better fit.  In the intermediate regime, we take a linear
interpolation between the two models.  These models do not
include the deuterium channel ($pp \rightarrow D\pi^+$),
important for low-energy collisions, and we include this separately
(note that this channel can be treated exactly). We assume that $\mu^+$ are 
produced isotropically in the $\pi^+$ decay, and we include
polarization in the subsequent $\mu^+$ decay (see \citealt{mosk} for
details).  Note that $F_+(\gamma,\gamma_\pi)$ can be found exactly,
unlike the pion distribution function.

Our treatment neglects two effects in pion production.  First, we ignore
the kaon channel, in which the collision forms a $K^+$ that subsequently
decays into either a $\pi^+$ or directly into a $\mu^+$.  \citet{miniati}
shows that the kaon channel provides only a small correction to the
dominant pion channel except at very high secondary energies (where the
total number of particles is very small regardless).  Second, we assume
that the ICM is composed entirely of hydrogen, while in reality $\sim 25\%$
of the mass is contained in helium nuclei.  \citet{mannheim} show that
modifying the ambient medium to a normal interstellar medium composition
increases the $\pi$ production rate by $\la 30 \%$.

The production rate depends critically on the input flux of cosmic ray
protons 
\bq
J_p(\gamma_p) = \frac{c}{4 \pi} \beta(\gamma_p) N_p(\gamma_p),
\label{eq:protonflux}
\eq
where $N_p(\gamma_p)$ is the differential number density of cosmic
ray protons with Lorentz factors between $(\gamma_p, \gamma_p +
d\gamma_p)$.  The protons are expected to be accelerated either by the
accretion shock surrounding the cluster or by shocks generated during
merger events.  Such shocks are collisionless and non-relativistic,
and hence could efficiently accelerate particles to relativistic energies
\citep{bell,bland-ost,bland-eichler}, as observed locally in supernova
remnants \citep{koyama95,koyama97,tanimori,muraishi,but}.  

We normalize the proton number density by requiring that the kinetic
energy density in relativistic protons be a fraction $\xi_{\rm CR}$ of the
total thermal energy density $u_{\rm th}=3n_e k_BT$ of the cluster: 
\bq
m_p c^2 \int (\gamma_p-1) N_p(\gamma_p) d\gamma_p = \xi_{\rm CR}
u_{\rm th}. 
\label{eq:pionnorm}
\eq
Shock acceleration models predict that the
distribution function of relativistic particles is $\propto p^{-s}$, where
$p$ is the particle momentum, $s = (r+2)/(r-1)$ and $r$ is the shock
compression ratio \citep{bland-eichler}.  The differential number density
of accelerated protons may therefore be written as
\bq 
N_p(\gamma_p) = N_{p,\,0} \frac{ \gamma_p
}{(\gamma_p^2 - 1)^{(s+1)/2}} \qquad \gamma_{p,\,{\rm min}} < \gamma_p <
\gamma_{p,\,{\rm max}},
\label{eq:protondf}
\eq
where $\gamma_{p,\, {\rm min}}$ and $\gamma_{p,\,{\rm max}}$ are the
minimum and maximum Lorentz factor to which a proton can be
accelerated in the cluster shocks.  For strong shocks in a gas with
adiabatic index $\Gamma=5/3$, $s = 2$.  \citet{miniati-prot} argue
that the accelerating shocks in cluster environments are relatively
weak, with Mach numbers $\la 4$--$5$, corresponding to power law
indices $s \ga 2.25$.  We consider the two cases $s=2$ and $s=3$ in
the following; these should bracket the (uncertain) distribution in
cluster environments.  The normalization constant $N_{p,\,0}$ is 
\bq
N_{p,\,0} = \frac{ 3.2 \times 10^{-7}}{ C^{\rm CR}_s } \, n_e T_{\rm
keV} \left( \frac{\xi_{\rm CR}}{0.1} \right) \cmden,
\label{eq:np0defn}
\eq
where $C^{\rm CR}_s$ depends on the power-law index of the cosmic ray
distribution function through equation (\ref{eq:pionnorm}); for
example, 
\bq
C^{\rm CR}_2 = 
\ln \left( \frac{\gamma_{p,\, {\rm max}} + \sqrt{\gamma_{p,\, {\rm
max}}^2 - 1}}{\gamma_{p,\, {\rm min}} + \sqrt{\gamma_{p,\, {\rm
min}}^2 - 1}} \right) + 
\frac{\sqrt{\gamma_{p,\, {\rm min}}^2 - 1}}{\gamma_{p,\, {\rm min}} +
1}  - \frac{\sqrt{\gamma_{p,\, {\rm max}}^2 - 1}}{\gamma_{p,\, {\rm
max}} + 1}
\label{eq:c2defn}
\eq
while
\bq
C^{\rm CR}_3 = \frac{1}{4} \left\{ \ln \left[
\frac{(\gamma_{p,\, {\rm 
max}} - 1)(\gamma_{p,\, {\rm min}} + 1)}{(\gamma_{p,\, {\rm
max}} + 1)(\gamma_{p,\, {\rm min}} - 1)} \right] + 
\frac{2}{\gamma_{p,\, {\rm min}} + 1} -
\frac{2}{\gamma_{p,\, {\rm max}} + 1} \right\}.
\label{eq:c3defn}
\eq 
Our fiducial value of $\xi_{\rm CR} = 0.1$ is probably a lower limit to the
energy deposited in cosmic ray protons \citep{fields,but}.  

The maximum energy to which protons may be accelerated is relatively well
determined.  Two factors limit $\gamma_{p,\,{\rm max}}$: the acceleration
time, $t_{\rm acc} \sim r_L c/v_s^2$, must be smaller than both the age of
the cluster and the cooling timescale; here $v_s$ is the shock velocity
(assumed to be close to the sound speed in the cluster) and $r_L$ is the
proton Larmor radius.  The cooling time of the protons (which is determined
primarily by pion production) exceeds the age of the universe, so the first
condition is the critical one, yielding $\gamma_{p,\,{\rm max}} \sim 5
\times 10^9 B_{\mu{\rm G}} T_{\rm keV}$, where $B_{\mu{\rm G}}$ is the
shock magnetic field in microgauss. Note that because $\gamma_{p,\,{\rm
max}} \gg 1$, $C_3$ is essentially independent of its precise value, while
it enters $C_2$ only logarithmically.

The value of $\gamma_{p,\,{\rm min}}$ is less certain.  Most analytic
models of diffusive shock acceleration 
assume that the particles retain a Maxwell-Boltzmann distribution below the
minimum injection momentum $p_{\rm min}$ and a (broken) power law above
that value (e.g., \citealt{kang,ellison}).  Normally $p_{\rm min}$ is a few
times the mean thermal momentum behind the accelerating shock, i.e. $p_{\rm
min} \approx 2 c_1 (m_p k_B T_{\rm sh})^{1/2}$, where $T_{\rm sh}$ is the
postshock temperature.  In principle, $c_1 \sim 2$--$3$ is determined by
the assumed energy density of the cosmic ray component.  However, because
$p_{\rm min}$ is located in the exponential tail of the Maxwellian
distribution, it depends only very weakly on the cosmic ray energy
density.  We therefore fix $c_1 \sim 2.3$, a value inferred from numerical
simulations of the shock acceleration process \citep{gieseler}. Then 
\bq
\gamma_{p,\,{\rm min}} - 1 \approx 3.4 \times 10^{-5} \, T_{\rm keV} \left(
\frac{c_1}{2.3} \right)^2 \left( \frac{\mach}{4} \right)^2,
\label{eq:gpmin}
\eq
where $\mach$ is the mean Mach number of the shocks.  In our
calculations, we fix $\mach=4$, as suggested by \citet{miniati-prot}.
Note that, particularly in the $s=3$ case, a non-negligible fraction
of the cosmic ray energy may be carried by protons with $\gamma_p <
\gamma_{\rm th}$.

The positron production rate via proton collisions $Q_{pp}(\gamma)$ is
shown in Figure \ref{fig:qpp} for input proton spectra with $s=2$ (solid
curves) and with $s=3$ (dotted curves).  Because $Q_{pp}$ is strictly
proportional to $\xi_{\rm CR} n_e^2$, we normalize our results by this
quantity.  The dependence on $T_e$ is slightly more subtle, so we show
results for $k_B T_e = 1,\,5,$ and $10 \keV$, from bottom to top.  The
ambient temperature does not affect the shape of the positron spectrum
(because in all cases $\gamma_{p,\,{\rm min}} < \gamma_{\rm th}$), but it
does affect the overall normalization factor $N_{p,\,0}$
[eq. (\ref{eq:np0defn})] and the \emph{fraction} of cosmic rays with
$\gamma_p > \gamma_{\rm th}$ [eq. (\ref{eq:gpmin})].  Nevertheless, our
results differ only slightly from the $Q_{pp} \propto T_e$ scaling that one
would naively expect from equation (\ref{eq:np0defn}).  Figure
\ref{fig:qpp} shows that the positrons are produced with a characteristic
$\gamma_+ \sim 200$; this indicates that $\ga 99\%$ of the energy deposited
in positrons will be lost to cooling before annihilation.  At high positron
energies, $Q_{pp} \propto \gamma^{-s}$.  This occurs because
$\sigma_{\pi^+}$ is only a weak function of proton energy for $\gamma_p \gg
1$.

For reference, the total positron production rate via proton
collisions is $Q_{pp} \approx
4.5 \times 10^{-17} n_e N_{p,\,0} \cmdensec$ for an $s=2$ cosmic ray
spectrum and $Q_{pp} \approx 1.5 \times 10^{-17} n_e N_{p,\,0}
\cmdensec$ for an $s=3$ spectrum. 

\subsubsection{ Other Positron Production Mechanisms }
\label{posprodother}

Cosmic ray protons can create positrons through two additional
channels:  photopair production ($p + \gamma \rightarrow p + e^+ + 
e^-$) and photopion production ($p + \gamma \rightarrow n + \pi^+$)
followed by $\pi^+$ decay.  In the diffuse IGM, these processes are
unimportant except for the highest energy cosmic rays because photopair
(or photopion) production requires $\gamma_p \ga 7 \times 10^8$ (or
$\gamma_p \ga 10^{11}$) if the photon originates from the CMB.
However, the high-temperature radiation field in clusters lowers these
thresholds dramatically to  $\gamma_p \ga 511 T_{\rm keV}^{-1}$ for
photopair production and $\gamma_p \ga 8 \times 10^4 T_{\rm keV}^{-1}$
for photopion production.  These channels could therefore in principle
play an important role in positron production in galaxy clusters.
The effects of the lowered thresholds have not to our
knowledge been addressed explicitly in the literature, so we will
estimate the total production rates here.  
We assume that the ambient
photon field arises from bremsstrahlung radiation, with a differential
energy density $u_\nu = 4 \pi r_c \varepsilon_\nu^{ff}/3 c$, where
$r_c$ is the core radius of the cluster and $\varepsilon_\nu^{ff}$ is
the bremsstrahlung spectral emissivity function \citep{rybicki}.

For either of these processes, the particle creation rate for a proton
traveling with Lorentz factor $\gamma_p$ may be written as \citep{mannheim}
\bq 
\frac{d n_+}{d t} = \frac{c}{2 \gamma_p^2} \int_{x'_{\rm th}/2
\gamma_p}^\infty\frac{d x}{x^2} \, \frac{d n}{d x} \int_{x'_{\rm th}}^{2
\gamma_p x} d x' \, x' \, \sigma(x'),
\label{eq:photcreaterate}
\eq 
where $dn/dx$ is the (isotropic) photon distribution function, $\sigma$
is the relevant collision cross-section, and $x'_{\rm th}$ is the minimum
photon energy in the proton rest frame needed to create the specified
particle(s).  If we approximate $\sigma$ as a constant and the exponential
cutoff in the bremsstrahlung photon spectrum as a step function with cutoff
$k_B T_e$, then the integrals may be easily evaluated. 
We find that the positron creation rate due to a single proton is 
\bq \frac{d
n_+}{d t} = K c \sigma \left[ \ln \left( \frac{\gamma_p}{\gamma_\ast}
\right) + \frac{\gamma_\ast^2}{\gamma_p^2} - 1 \right],
\label{eq:photloss}
\eq
where $K$ is a constant evaluated below and $\gamma_\ast = x_{\rm
th}'/2 k_B T_e$ is the minimum proton Lorentz factor required to
produce the particle(s) given our assumed photon spectrum.  

If we assume a cosmic ray spectrum $N_p(\gamma_p) = N_{p,\,0}
\gamma_p^{-2}$, maximizing the effects of protons with $\gamma_p >
\gamma_\ast$, the total rate of positron production $Q$ is
\bq
Q = \frac{ K c \sigma N_{p,\,0}}{3 \gamma_\ast}.
\label{eq:photsource}
\eq

For the photoproduction processes we consider, $\sigma_{\gamma e^+e^-} \sim
6 \mbarn$ \citep{chod}, while $\sigma_{\gamma \pi^+} \sim 150 \mubarn$
\citep{bhatt}.  Evaluating $K$ based on the assumed energy density of the
cluster, we then find \bq Q_{\gamma e^+e^-} \sim 3 \times 10^{-26}
N_{p,\,0} T_{\rm keV}^{1/2} \left( \frac{n_e}{10^{-3} \cmden} \right)^2
\left( \frac{r_c}{200 \kpc} \right) \cmden
\label{eq:qphotopair}
\eq
and
\bq
Q_{\gamma \pi^+} \sim 5 \times 10^{-30} N_{p,\,0} T_{\rm keV}^{1/2}
\left( \frac{n_e}{10^{-3} \cmden} \right)^2 
\left( \frac{r_c}{200 \kpc} \right) \cmden.
\label{eq:qphotopion}
\eq

Thus, $Q_{\gamma \pi} \ll Q_{\gamma e^+ e^-} \ll Q_{pp}$, and the two
photo-collision processes may indeed be neglected in even the most luminous
clusters.  (We find this conclusion to hold regardless of the detailed
photon spectrum assumed, provided that the total energy density is fixed.)
Note that, despite our conclusion that $Q_{\gamma \pi} \ll Q_{\gamma
e^+ e^-}$, photopion production is a more significant energy loss mechanism 
for high-energy protons than is photopair production.  The difference in
significance results from cosmic ray protons with $511 \la
\gamma_p T_{\rm keV} \la 8 \times 10^4$ being energetic enough to produce
pairs (but not pions) and from the greater inelasticity of photopion
production as compared to photopair production \citep{mannheim}.

\section{ Calculation of the Positron Distribution Function }
\label{posdfsoln}

\subsection{ Positron Injection by an AGN }
\label{agnsoln}

For positrons which are injected impulsively by a short-lived AGN
embedded in the cluster, the source term $Q(\gamma) = 0$ in equation
(\ref{eq:dfevol}).  If we further neglect annihilation of the cooling
positrons, the calculation of the positron distribution function is
straightforward \citep{sarazin}.  This approximation is valid because for
any $\gamma > \gamma_{\rm eq}$, the annihilation time is much longer than
the cooling time.  (See \S \ref{annspecagn} for further discussion of this
point.)

The calculation proceeds as follows.  We first assume that the AGN
injects positrons into the cluster with a distribution function
$N_+(\gamma,t_i)$, where $t_i$ is the injection time.  Consider a
positron with Lorentz factor $\gamma_i$ at time $t_i$.  At a later
time $t$, the positron will have cooled to a Lorentz factor
$\gamma_0$.  Number conservation demands that 
\bq
\int_{\gamma_0}^\infty N_+(\gamma',t) d\gamma' = 
\int_{\gamma_i}^\infty N_+(\gamma',t_i) d\gamma'.
\label{eq:numcon}
\eq
Differentiating, we find \citep{sarazin}
\bq
N_+(\gamma_0,t) = N_+(\gamma_i,t_i) \left. \frac{d \gamma_i}{d
\gamma_0} \right|_t ,
\label{eq:agndf}
\eq
with $\gamma_i$ computed using the cooling function $b(\gamma,t)$ given
in \S 2.1.

At any time $t > t_i$, the number of positrons that have not yet cooled
$n_u(t)$ may be calculated by integrating the distribution function
over all Lorentz factors $\gamma > \gamma_{\rm eq}$.  The number
density of thermalized positrons $n_+(t)$ then evolves according to
\bq
\frac{d n_+}{d t} = - \frac{d n_u}{d t}  - n_+ A(\gamma_{\rm eq}),
\label{eq:ncoolagnevol}
\eq
where the first term is the rate at which positrons thermalize and the
second term is the rate at which they annihilate.  Thus
\bq
n_+(t) = e^{-A(\gamma_{\rm eq}) t} \int^t_{t_i} d t' e^{A(\gamma_{\rm
eq}) t'} \left| \frac{d n_u}{d t} \right|.
\label{eq:ncoolagnsoln}
\eq

We calculate the initial distribution function $N_+(\gamma,t_i)$
by expressing the total positron energy density (including the rest
mass) immediately after injection as 
\bq
u_{\rm rel} = m_e c^2 \int \gamma N_+(\gamma,t_i) d \gamma = 
\xi_{\rm AGN} \frac{L_K \tau}{V},
\label{eq:urelagn}
\eq 
where $L_K$ is the total mechanical luminosity of the AGN, $\tau$ is
the lifetime of the AGN (in a single duty cycle), $V$ is the volume over
which the relativistic particles are mixed, and $\xi_{\rm AGN}$ is the
fraction of the injected energy in the positron component (see below).  For
typical parameter values, the energy input from a central AGN is
significant compared to the thermal energy stored in the cluster core and
could, in some cases, balance the energy lost to cooling flow radiation
\citep{bohringer}.  If we assume that the positrons initially have a power
law distribution in momentum, 
\bq 
N_+(\gamma,t_i) = n_i \frac{ \gamma
}{(\gamma^2 - 1)^{(s+1)/2}} \qquad \gamma_{+,\,{\rm min}} < \gamma <
\gamma_{+,\,{\rm max}},
\label{eq:posdfi}
\eq
the normalization constant $n_i$ is 
\bq
n_i = \frac{3.9 \times 10^{-7}}{ C^{\rm AGN}_s } \left( \frac{\xi_{\rm
AGN}}{0.1} \right) 
\left( \frac{L_K}{10^{45} \ergs} \right) \left( \frac{\tau}{10^8 \yr}
\right) \left( \frac{r_{\rm mix}}{200 \kpc} \right)^{-3} \cmden,
\label{eq:ni}
\eq
where $r_{\rm mix} \sim r_c$ is the radius out to which positrons are
distributed and $C^{\rm AGN}_s$ depends on the power-law index of the
distribution and is determined by equation (\ref{eq:urelagn}):
\bq
C^{\rm AGN}_2 = 
\ln \left( \frac{\gamma_{+,\, {\rm max}} + \sqrt{\gamma_{+,\, {\rm
max}}^2 - 1}}{\gamma_{+,\, {\rm min}} + \sqrt{\gamma_{+,\, {\rm
min}}^2 - 1}} \right) + 
\frac{\gamma_{+,\, {\rm min}}}{\sqrt{\gamma_{+,\, {\rm min}}^2 - 1}}
- \frac{\gamma_{+,\, {\rm max}}}{\sqrt{\gamma_{+,\, {\rm max}}^2 - 1}}
\label{eq:c2agndefn}
\eq
and
\bq
C^{\rm AGN}_3 = \frac{1}{4} \left\{ \ln \left[
\frac{(\gamma_{+,\, {\rm 
max}} - 1)(\gamma_{+,\, {\rm min}} + 1)}{(\gamma_{+,\, {\rm
max}} + 1)(\gamma_{+,\, {\rm min}} - 1)} \right] + 
\frac{\gamma_{+,\, {\rm min}}}{\gamma_{+,\,{\rm min}}^2-1} -
\frac{\gamma_{+,\, {\rm max}}}{\gamma_{+,\,{\rm max}}^2-1} \right\}.
\label{eq:c3agndefn}
\eq

In equation (\ref{eq:posdfi}), $\gamma_{+,\,{\rm max}} \sim 4 \times 10^7
\mach (B_{sh,\,\mu{\rm G}} T_{\rm keV})^{1/2}$ is fixed by equating the
acceleration time to the IC cooling time.  $B_{sh}$ thus refers to the
magnetic field in the shocks inside the accelerating region of the jet
system; for simplicity, we take $B_{sh} = B$.  We note that the magnetic
field in the accelerating region may have considerably larger values than
the ambient cluster field because of the powerful magnetic fields in AGN
jets (e.g., \citealt{bbr}).  However, the dependence on $\gamma_{+,\,{\rm
max}}$ is only logarithmic and assuming a larger $B$ value introduces only
a small change to our results.  We also assume that $\mach \sim 10$, a
value appropriate for the hotspot velocity at near the end of the active
phase [see, for example, the model in \citet{bigm}].

As above, we determine $\gamma_{+,\,{\rm min}}$ assuming that the
minimum cosmic ray momentum is a few times the mean thermal momentum.
For positrons, this implies that 
\bq
\gamma_{+,\,{\rm min}}^2 - 1 = 0.74 \, T_{\rm keV} \left( \frac{c_1}{2.3}
\right)^2 \left( \frac{\mach}{10} \right)^2.
\label{eq:gplusmin}
\eq
Note that $(\gamma_{+,\,{\rm min}} - 1)$ is of order unity for
reasonable cluster parameters. 

The distribution function in equation (\ref{eq:posdfi}) should be
thought of as referring to the positron component once it eventually
escapes the entire region of interaction between the relativistic jets
and the ambient medium.  It is not necessarily the same distribution
the positrons have while inside the jet, where cooling may be
efficient \citep{bbr} but where the positrons are expected to have a
bulk Lorentz factor $\gamma_{\rm jet} \sim 10$.  As they travel along the 
jet and particularly when they pass through the hotspot, surviving
positrons will presumably be re-accelerated to a fresh distribution
given by equation (\ref{eq:posdfi}).  The normalization factor $\xi_{\rm
AGN}$ thus represents the net fraction of the interaction energy $L_K
\tau$ transferred to the escaping positrons.  Note that we assume that
all of the positrons are accelerated by the shocks.  If only a
fraction are actually accelerated, the total number of positrons (and
hence the annihilation signal) would only increase. 

\begin{figure}[t]
\plotone{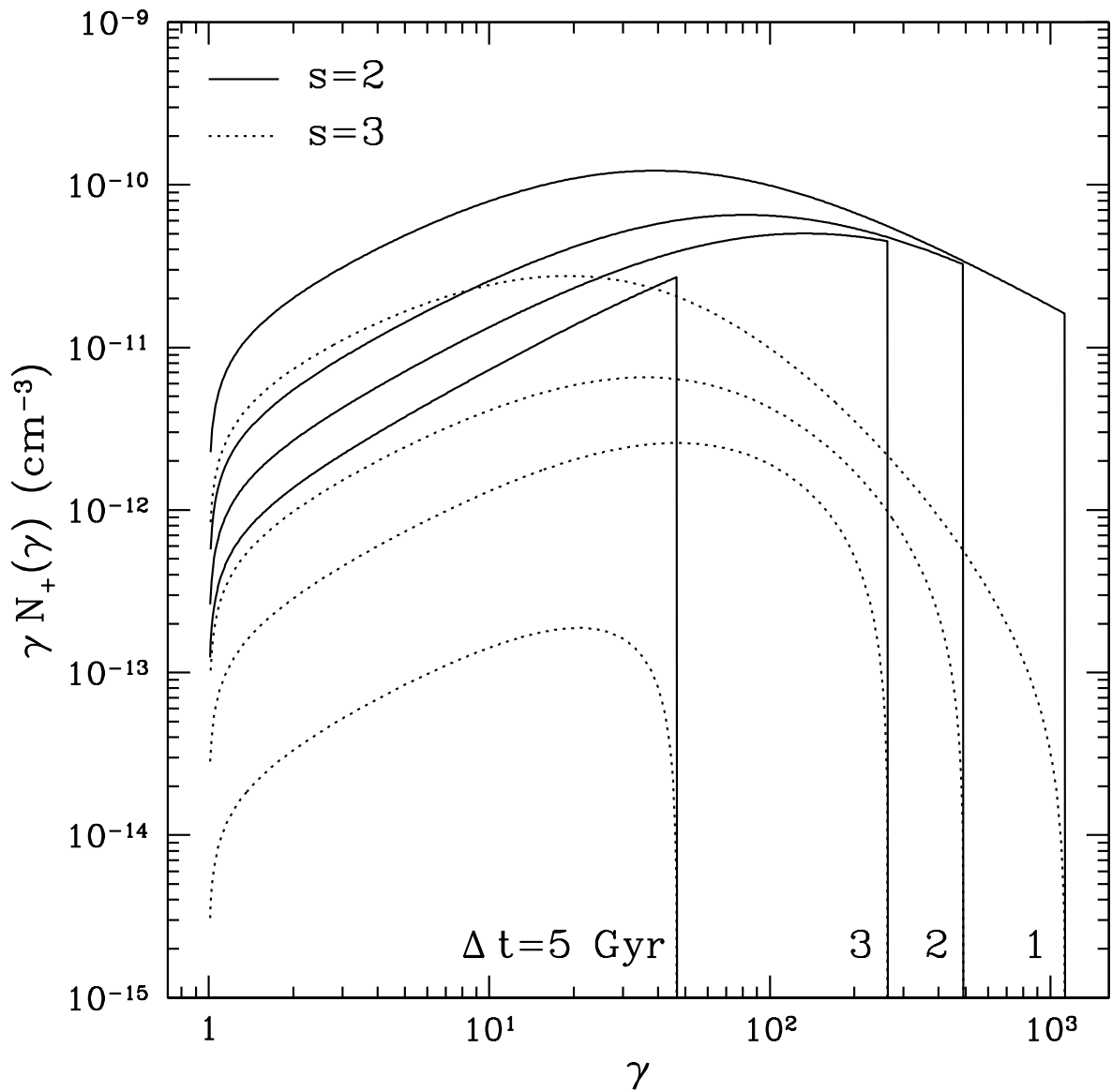}
\caption{ The distribution function of non-thermalized positrons
$N_+(\gamma,t)$ for injection via an AGN at various times.  Shown are
an $s=2$ injection spectrum (\emph{solid curves}) and an $s=3$
injection spectrum (\emph{dotted curves}).  In each case, curves
correspond to $\Delta t = t_0 -t_i = (1,\,2,\,3,\,5) \times 10^9 \yr$,
from top to bottom.  All cases assume standard cluster parameters (see
text). }
\label{fig:agndf}
\end{figure}

Figure \ref{fig:agndf} shows $N_+(\gamma,t)$ in the AGN scenario at
several different time intervals after injection.  We present results
for an $s=2$ injection spectrum (solid curves) and for an $s=3$
injection spectrum (dotted curves); in each case, curves show $t_0 -
t_i = (1,\, 2,\, 3,\, 5) \times 10^9 \yr$, from top to bottom.  Here
$t_0$ is the (present-day) age of the universe in our assumed
cosmology.  All cases assume what we will refer to as our ``standard
cluster parameters:'' $L_K = 10^{45} \ergs$, $\tau = 10^8 \yr$, $n_e =
10^{-3} \cmden$, $k_B T_e = 1 \keV$, $B = 3 \microgauss$, $r_{\rm mix}
= 200 \kpc$, and $\xi_{\rm AGN} = 0.1$.  Note that $N_+(\gamma,t) \propto 
\xi_{\rm AGN} L_K \tau r_{\rm mix}^{-3}$, with a more subtle
dependence on $n_e$ (see below).  This scaling holds as long as our
approximation of instantaneous positron injection is valid, namely for
$\tau \la 3 \times 10^8 \yr$.  The dependence on temperature enters
only through $\gamma_{+,\,{\rm min}}$.

The main feature in the distribution function is the sharp cutoff at a
Lorentz factor $\gamma_{\rm max}(t)$.  At any given time after
injection, this corresponds to 
\bq
\gamma_{\rm max}(t) = \lim_{\gamma_i \rightarrow \infty}
\gamma(\gamma_i,t,t_i), 
\label{eq:gmaxdefn}
\eq
where $\gamma(\gamma_i,t)$ is the Lorentz factor at time $t$ of a
positron injected with $\gamma_i$ at time $t_i$.  The limit is
well-defined because $b(\gamma) \propto \gamma^2$ for large Lorentz
factors; however, we do not solve for it in closed form here.
\citet{sarazin} gives expressions for $\gamma_{\rm max}(t)$ in some
simple limits. 

\begin{figure}[t]
\plotone{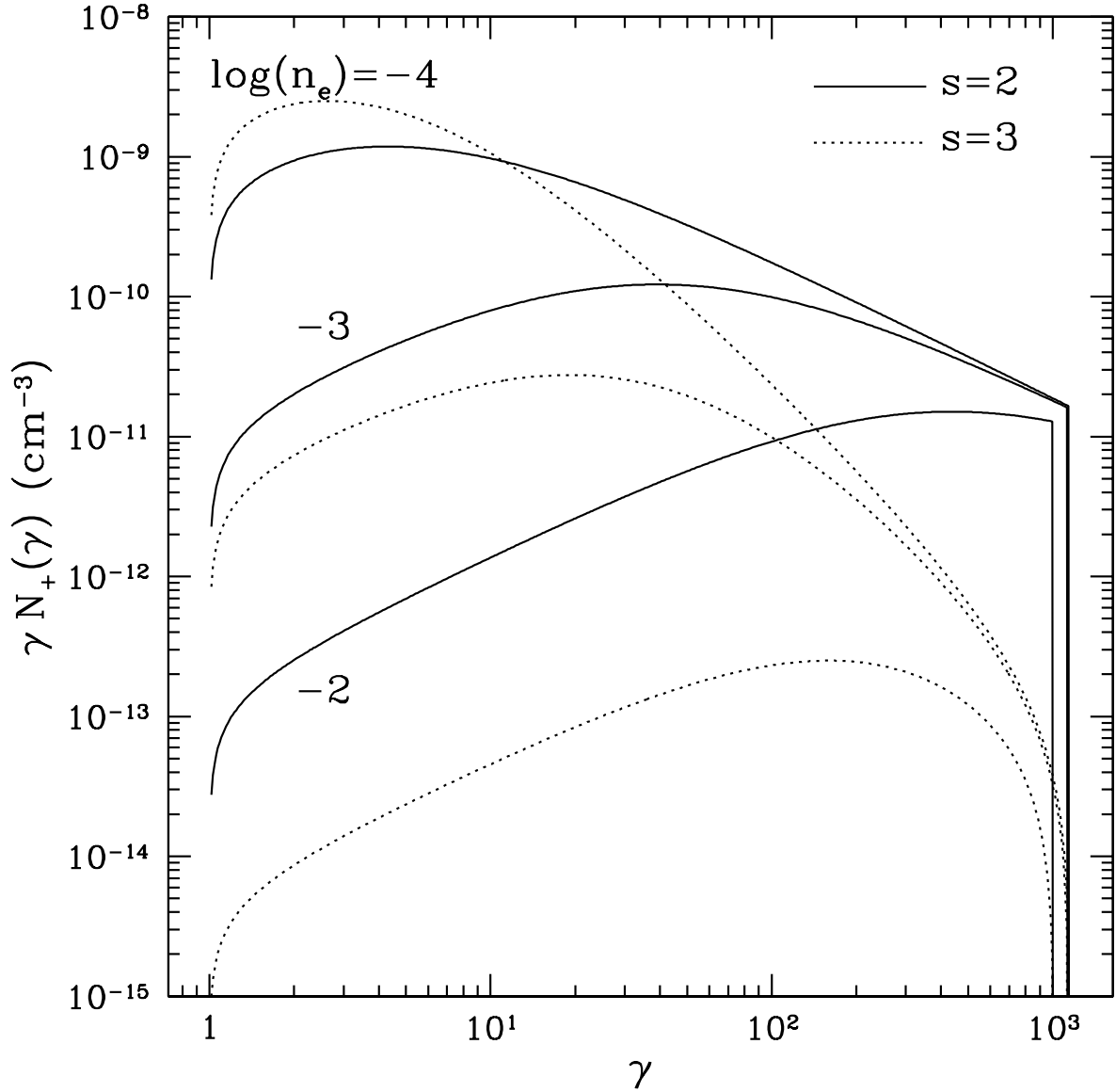}
\caption{ The distribution function of non-thermalized positrons
$N_+(\gamma,t)$ for injection via an AGN at various  ambient
densities.  Shown are an $s=2$ injection spectrum (\emph{solid
curves}) and an $s=3$ injection spectrum (\emph{dotted curves}).  In
each case, curves correspond to $n_e = 10^{-4},\, 10^{-3},$ and
$10^{-2} \cmden$, from bottom to top.  All cases assume $t_0 - t_i =
10^9 \yr$ and standard cluster parameters (with the exception of
$n_e$; see text). }
\label{fig:agndfden}
\end{figure}

Figure \ref{fig:agndfden} shows the effects of varying the cluster core
density $n_e$ on the distribution function of the uncooled
positrons. Again, the solid (dotted) curves show results for an $s=2$ $(3)$
injection spectrum.  All assume $t_0 - t_i = 10^9 \yr$ and, aside from the
density, standard cluster parameters.  From top to bottom, the curves in
each case assume $n_e = 10^{-4},\, 10^{-3},$ and $10^{-2} \cmden$.  Perhaps
surprisingly, $N_+(\gamma)$ decreases as $n_e$ increases.  This
non-intuitive behavior is a result of the cooling rates.  Above the cooling
``bottleneck'' at $\gamma \sim 200(n_e/10^{-3} \cmden)^{1/2}$, the cooling
time is independent of density [see equation (\ref{eq:bsynic2})]; below
this Lorentz factor, the cooling time is inversely proportional to density
[see equation (\ref{eq:bcoul2})].  In other words, once positrons
enter the Coulomb cooling regime, they thermalize at a rate
proportional to $n_e$.  Thus, near 
the bottleneck $N_+(\gamma)$ is independent of density, while below it
$N_+(\gamma)$ actually decreases with increasing density because positrons
enter the thermalized component faster.  (This density dependence does
\emph{not} hold for the thermalized positron density $n_+$.)

The evolution of $n_+$ as a function of cosmic time is shown in the
bottom panels of Figures \ref{fig:linetimes2} and \ref{fig:linetimes3}.
Figure \ref{fig:linetimes2} shows results for an $s=2$ positron
injection spectrum; the solid curves assume $n_e = 10^{-3} \cmden$
with $t_0 - t_i = 1,\,2,\,3,$ and $5 \times 10^9 \yr$, while the
dotted curve assumes $n_e = 10^{-4} \cmden$ with $t_0 - t_i = 5 \times
10^9 \yr$ and the dashed curve assumes $n_e = 10^{-2} \cmden$ with
$t_0 - t_i = 5 \times 10^9 \yr$.  Figure \ref{fig:linetimes3} shows a
similar set of results for an $s=3$ injection spectrum.  In all
cases, we assume standard cluster parameters (except for the electron
density).  We find that steepening the power law slope $s$ increases
the number of thermalized positrons because the total number of
positrons produced by the AGN increases as $s$ increases.

At very early times after injection, the thermalization rate exceeds the
annihilation rate and $n_+$ rises rapidly, particularly for a steep
injection spectrum in which the number density of particles is heavily
weighted toward small $\gamma$.  However, once the low-$\gamma$ positrons
cool, the thermalization rate falls rapidly with time (because fewer
positrons begin with large Lorentz factors) and annihilation begins to
dominate.  Therefore, at late times $n_+ \propto e^{-A(\gamma_{\rm
eq})(t-t_i)}$.

The dependence on density is more complex.  The total number density of
injected positrons (and hence the maximum value of $n_+$) is independent of
$n_e$, and so is the peak value of $n_+$. However, because $b_{\rm Coul}
\propto n_e$, the value of $n_+$ peaks later for smaller ambient densities.
Furthermore, the annihilation timescale is $\tau_{\rm ann} \sim
A(\gamma_{\rm eq})^{-1} \sim 4 \times 10^9 (n_e/10^{-3} \cmden)^{-1} \yr$;
for $n_e = 10^{-4} \cmden$, $\tau_{\rm ann} \ga H_0^{-1}$ and $n_+$ does
not change substantially over the lifetime of the cluster.

Note that the distribution functions derived in this section depend
weakly on the cosmological redshift through the corresponding
dependence of the IC cooling rate.

\subsection{ Steady Injection by Cosmic Rays }
\label{crsoln}

In the case of steady injection, with $Q_{pp}(\gamma)$ computed as in \S
\ref{ppcoll}, we can calculate the positron distribution function reached
at steady-state (i.e., when the total production rate of positrons equals
the annihilation rate).  In this case, all quantities in equation
(\ref{eq:dfevol}) are independent of time and the distribution function of
uncooled positrons is
\bq
N_+(\gamma) = \frac{1}{\mu(\gamma)} \int_{\gamma}^\infty d \gamma'
\mu(\gamma') \frac{Q(\gamma')}{b(\gamma')},
\label{eq:steadydf}
\eq
where 
\bq
\mu(\gamma) = \exp \int_{\gamma_1}^{\gamma} \frac{d \gamma'}{b(\gamma')}
\left[ \frac{d b}{d \gamma'} - A(\gamma') \right]
\label{eq:mudefn}
\eq
and $\gamma_1$ is an arbitrary constant.  Note that, if we neglect
annihilation, this reduces to the equilibrium distribution found by
\citet{sarazin}. The number density of thermalized positrons may then
be found by requiring the total number density of positrons to be
independent of time: 
\bq
n_+ = \frac{1}{A(\gamma_{\rm eq})} \int_{\gamma_{\rm eq}}^\infty 
d\gamma \, \left[ Q_{pp}(\gamma) - N_+(\gamma) A(\gamma) \right].
\label{eq:ncoolsteady}
\eq

Our scheme requires that enough time has elapsed since the cosmic ray
proton component first formed for the annihilations to reach
equilibrium with production.  Because the annihilation cross section
decreases rapidly with $\gamma_+$, this essentially requires that the
thermalized component has grown enough that $A(\gamma_{\rm eq}) n_+
\approx Q_{pp}$.  This requires a time $\tau_{\rm eq} \sim \tau_{\rm
ann}$.  We therefore expect equilibrium to be a satisfactory
assumption for $n_e \ga 3 \times 10^{-4} \cmden$, a condition 
well-statified by rich clusters.

\begin{figure}[t]
\plotone{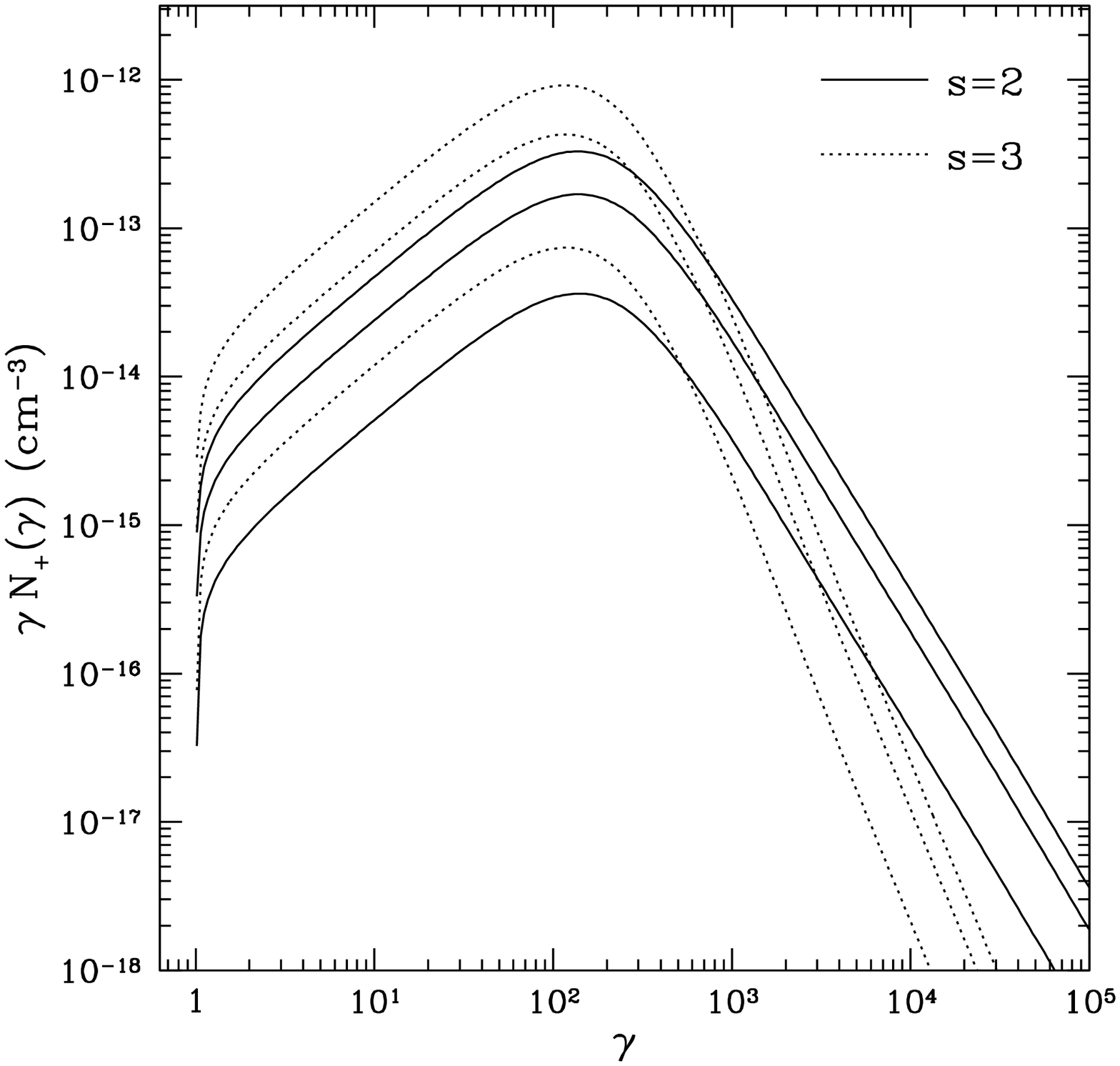}
\caption{ The equilibrium distribution function of non-thermalized
positrons $N_+(\gamma)$ assuming steady $Q_{pp}(\gamma)$, for
various cluster temperatures. Shown are an $s=2$ proton spectrum
(\emph{solid curves}) and an $s=3$ proton spectrum
(\emph{dotted curves}).  In each case, curves correspond to $k_B T_e =
1,\, 5,$ and $10 \keV$, from bottom to top.  All curves assume $n_e =
10^{-3} \cmden$, $B = 3 \microgauss$, and $\xi_{\rm CR} = 0.1$.  
}
\label{fig:piondf}
\end{figure}

Figure \ref{fig:piondf} shows the equilibrium positron distribution
$N_+(\gamma)$ with $Q_{pp}(\gamma)$ from \S \ref{ppcoll}.  We show results
for an $s=2$ cosmic ray proton spectrum (solid curves) and for an $s=3$
proton spectrum (dotted curves).  In each case, curves correspond to $k_B
T_e = 1,\, 5,$ and $10 \keV$, from bottom to top.  All assume $n_e =
10^{-3} \cmden$, $B = 3 \microgauss$, and $\xi_{\rm CR} = 0.1$.  The
distribution function is strictly proportional to $\xi_{\rm CR}$.  Although
$T_e$ does affect $\gamma_{p,\,{\rm min}}$ and hence the fraction of cosmic
ray protons able to produce pions, we find that $N_+(\gamma)$ is still very
nearly proportional to $T_e$.

\begin{figure}[t]
\plotone{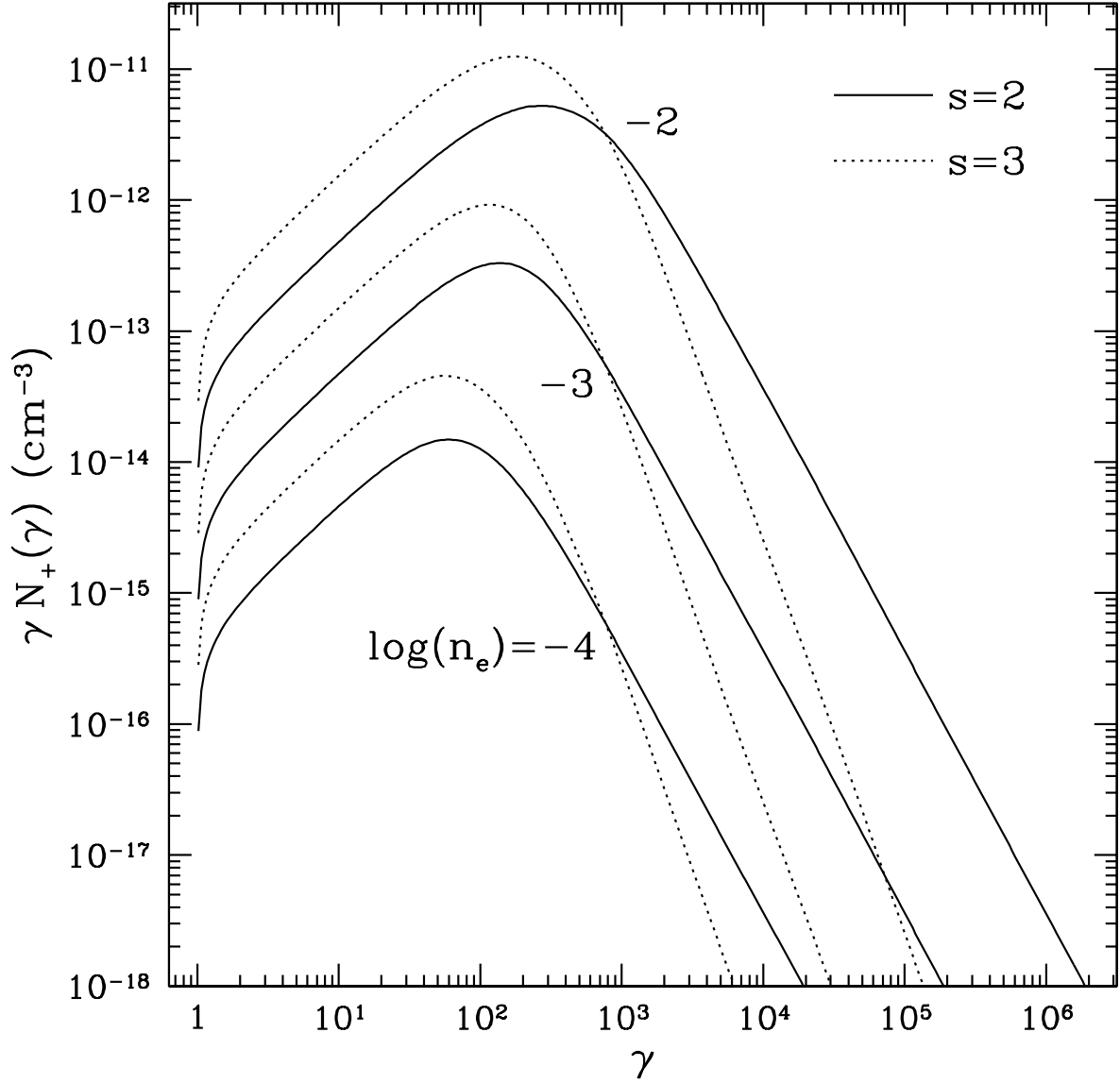}
\caption{ The equilibrium distribution function of non-thermalized
positrons $N_+(\gamma)$ assuming steady $Q_{pp}(\gamma)$, for
various cluster densities. Shown are an $s=2$ proton spectrum
(\emph{solid curves}) and an $s=3$ proton spectrum
(\emph{dotted curves}).  In each case, curves correspond to $n_e =
10^{-4},\, 10^{-3},$ and $10^{-2} \cmden$, from bottom to top. All
curves assume $k_B T_e = 10 \keV$, $B = 3 \microgauss$, and $\xi_{\rm
CR} = 0.1$.   } 
\label{fig:piondfden}
\end{figure}

Figure \ref{fig:piondfden} illustrates how $N_+(\gamma)$ depends
on the density of the ambient medium.  As before, the solid (dotted)
curves show results for an $s=2$ $(3)$ proton spectrum.  The cluster
is assumed to have $\xi_{\rm CR} = 0.1$, $B = 3 \microgauss$, and $k_B T_e =
10 \keV$, with $n_e = 10^{-4},\, 10^{-3},$ and $10^{-2} \cmden$, from
bottom to top.  We see that the peak in the distribution function
moves to larger Lorentz factors as density increases and Coulomb
cooling becomes more efficient.  Note, however, that the cluster with
$n_e = 10^{-4} \cmden$ may not yet have reached steady-state.

In all cases, $N_+(\gamma) \propto \gamma^{-(s+1)}$ at high energies
where synchrotron/IC cooling dominates and annihilation can be 
neglected \citep{sarazin}.  The distribution function peaks where the
efficiency of both synchrotron/IC and Coulomb cooling is low ($\gamma
\sim 100$) and then falls again at small Lorentz factors as Coulomb
cooling takes over. 

The equilibrium density of thermalized positrons $n_+$ is not shown in
Figures \ref{fig:piondf} and \ref{fig:piondfden} because it scales
simply with the cluster parameters:
\bq
n_+ \sim 8 \times 10^{-14} X_s^{\rm CR} \, T_{\rm keV} \left(
\frac{n_e}{10^{-3} \cmden} \right) \left( \frac{\xi_{\rm CR}}{0.1}
\right) \cmden,
\label{eq:npluspion}
\eq
where $X_s^{\rm CR} = 1$ ($2.5$) for an $s=2$ ($s=3$) proton
spectrum. 

\section{ The Annihilation Spectrum } 
\label{annspec}

In the following subsections we calculate the emissivity of
annihilation photons from the ICM using equations
(\ref{eq:posannspec}) and (\ref{eq:posanntherm}) for both the AGN (\S
\ref{annspecagn}) and the steady injection (\S \ref{pionspec})
scenarios.  In each case, we use the distribution function
$N_+(\gamma)$ found in the appropriate subsection of \S \ref{posdfsoln}. 

\subsection{ Positron Injection by an AGN }
\label{annspecagn}

For injection by an AGN, we ignored annihilation of non-thermalized
positrons when calculating $N_+(\gamma,t)$, although we did include
annihilation of the thermalized component.  Here we calculate the
instantaneous emissivity expected for the resulting distribution.

\begin{figure}[t]
\plotone{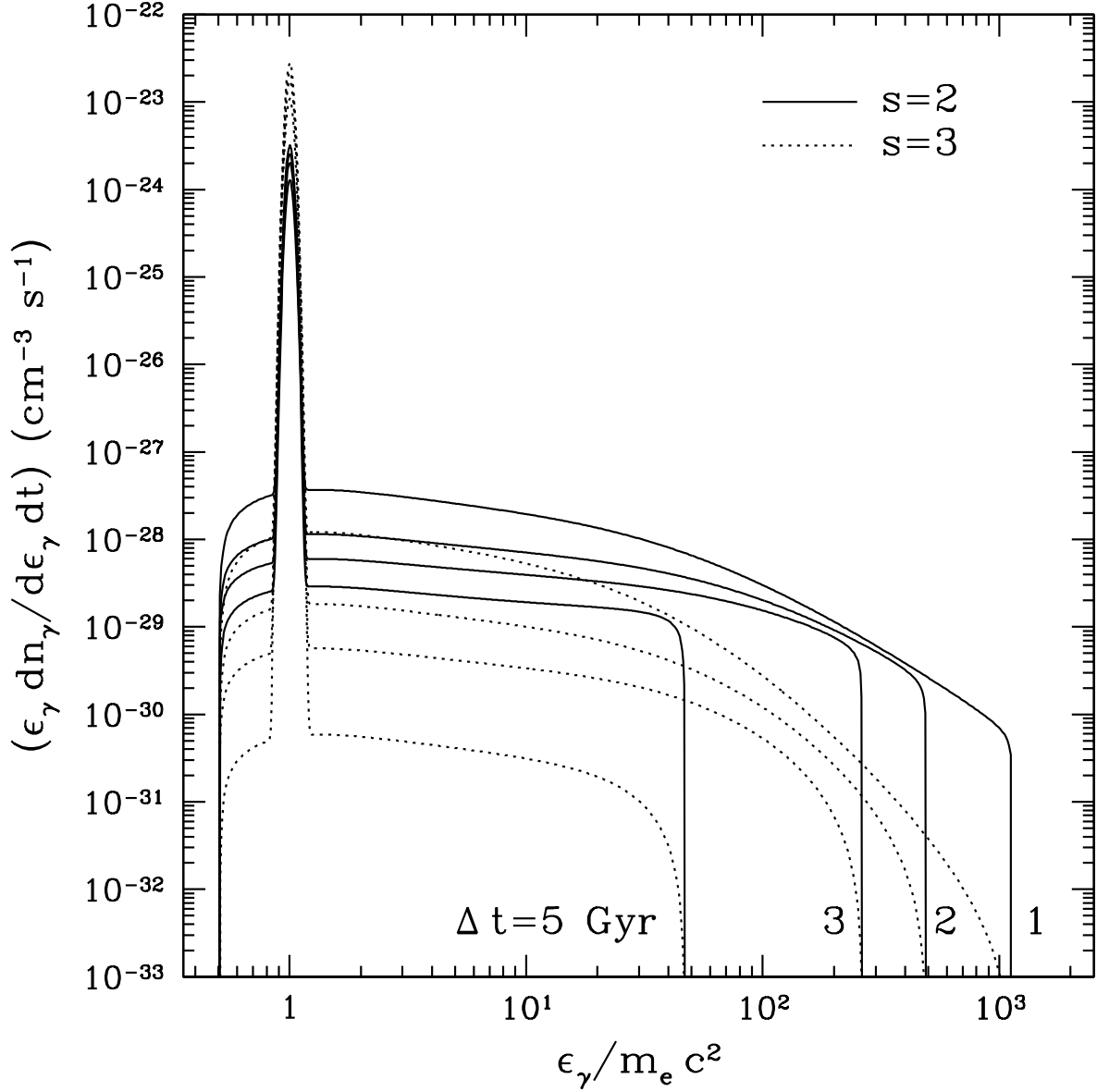}
\caption{ The annihilation emissivity $(d n_\gamma/d\epsilon_\gamma\,dt)$
for injection via an AGN at various times.  Shown are an $s=2$
injection spectrum (\emph{solid curves}) and an $s=3$ injection spectrum
(\emph{dotted curves}).  In each case, curves correspond to $\Delta t
= t_0 -t_i = (1,\,2,\,3,\,5) \times 10^9 \yr$, from top to bottom in
continuum level.  All curves assume standard cluster parameters (see
text). }
\label{fig:dndkagn}
\end{figure}

Figure \ref{fig:dndkagn} shows the annihilation emissivity $(d
n_\gamma/d \epsilon_\gamma d t)$ for the same cases as in
Figure \ref{fig:agndf}.  The continuum, in which the emissivity is
approximately $\propto \epsilon_\gamma^{-1}$, is generated by the
uncooled positrons, while the line is generated by the thermalized
positrons.  The sharp cutoffs in the continuum occur because of our
{\it cosmic ray approximation} in which $\gamma_- = 1$.  In this
approximation, only photons with $[\gamma_{+,\,{\rm
max}}(1-\beta_{+,\,{\rm max}}) + 1]/2 \leq 
\epsilon_\gamma/m_e c^2 \leq [\gamma_{+,\,{\rm 
max}}(1+\beta_{+,\,{\rm max}}) + 1]/2$ can be generated
\citep{svensson}; in reality, the non-zero high-energy tails of the
ambient electron distribution (and collisions with relativistic cosmic
ray electrons) will cause the cutoffs to be somewhat smoother.  The
time evolution of line emissivity is discussed in detail below. 

Figure \ref{fig:dndkagn} justifies our approach of neglecting
annihilation in calculating the distribution function.  Consider, for
example, the case in which $t_0 - t_i = 10^9 \yr$ for an $s=2$
injection spectrum.  The total annihilation emissivity in the
continuum is $\sim 5 \times 10^{-28} \cmdensec$, yielding a total
number of annihilations of non-thermalized positrons since injection
of $\sim 8 \times 10^{-12} \cmden$.  We calculate the total density of
non-thermalized positrons to be $\sim 2 \times 10^{-10} \cmden$,
indicating that only $\sim 5\%$ of the positrons annihilate before
cooling. 

\begin{figure}[t]
\plotone{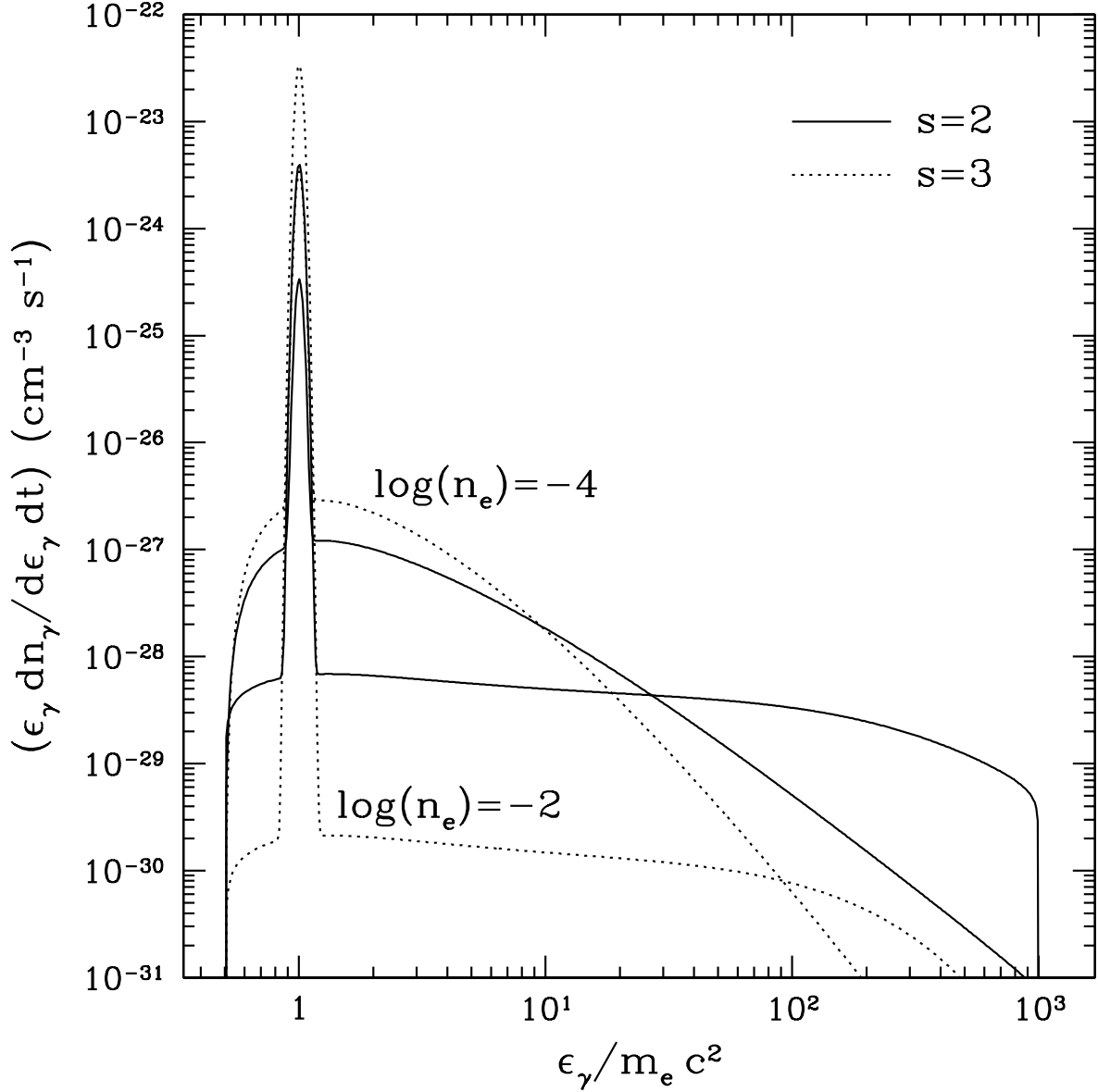}
\caption{ The annihilation emissivity $(d n_\gamma/d\epsilon_\gamma\,dt)$
for injection via an AGN at various ambient densities.  Shown are an
$s=2$ injection spectrum (\emph{solid curves}) and an $s=3$ injection
spectrum (\emph{dotted curves}).  In each case, we show results for $n_e =
10^{-4}$ and $10^{-2} \cmden$.  All curves assume standard cluster
parameters (with the exception of $n_e$; see text) and $t_0-t_i =
10^9 \yr$. } 
\label{fig:dndkagnden}
\end{figure}

The line emissivity at a given time is proportional to $ \xi_{\rm AGN} L_K
\tau r_{\rm mix}^{-3}$, and so the total luminosity of the cluster is
independent of $r_{\rm mix}$.  The dependence on ambient density is
somewhat more complicated.  This is displayed in Figure
\ref{fig:dndkagnden}, which shows the annihilation emissivity for clusters
with densities $n_e = 10^{-4} \cmden$ and $n_e = 10^{-2} \cmden$ for $s=2$
and $s=3$ positron input spectra (solid and dotted curves, respectively).
Standard cluster parameters (with the exception of density) are assumed,
with $t_0-t_i=10^9 \yr$.  The continuum emissivity depends on the shape of
$N_+(\gamma,t)$, which has a non-negligible dependence on $n_e$. Note that
more continuum photons are produced in \emph{lower} density media.  This is
because, as shown in Figure \ref{fig:agndfden}, a larger fraction of the
positrons have thermalized by this epoch in high density clusters, and the
thermalized component contributes only to the line emission.  The
dependence of the peak line luminosity on density is very different from
that of the continuum (see below).  The cluster temperature affects both
the normalization of the continuum (although only relatively weakly) and
the width of the annihilation line [the full width at half-maximum is
approximately $\sim 32 T_{\rm keV}^{1/2} \keV$; see equation
(\ref{eq:posanntherm})].  This temperature-dependent line broadening is
displayed graphically in Figure \ref{fig:dndkicn2} below.  

Observationally, the critical observable is the emissivity of the
annihilation line; this quantity varies strongly as a function of the
source age.  We show the evolution of the line emissivity $\dot{n}_{\rm
line}$ with time in the top panels of Figures \ref{fig:linetimes2} and
\ref{fig:linetimes3} for the same cases shown in the bottom panels.  The
emissivity is proportional to the number of thermalized positrons, so
at late times $\dot{n}_{\rm line} \propto n_e e^{-A(\gamma_{\rm
eq})(t-t_i)}$.  Thus the peak signal decreases as density decreases,
but it also fades with a timescale $\tau_{\rm ann} \propto
n_e^{-1}$. 

The emissivities calculated in this section depend weakly on redshift
through the IC cooling rate.

\subsection{ Steady Injection by Cosmic Rays }
\label{pionspec}

In the steady injection scenario, we have calculated $N_+(\gamma)$
self-consistently, including annihilation.  The resultant photon
emissivities are shown in Figures \ref{fig:dndkicn2} and
\ref{fig:dndkicn3} for $s=2$ and $s=3$ proton spectra, respectively.
In each case, the solid curves show the annihilation emissivity, with
$k_B T_e = 1 \keV$ (bottom) and $k_B T_e = 10 \keV$ (top).  

In these figures, we also show the inverse Compton emissivity
determined from the equilibrium positron spectrum $N_+(\gamma)$.  In
the energy range of interest ($\gamma \approx 10^4$), $N_+(\gamma)$
has assumed its asymptotic power law form, and the inverse Compton
spectrum can be estimated easily \citep{rybicki}:
\bq
\left. \epsilon_\gamma \frac{d n_\gamma}{d \epsilon_\gamma dt}
\right|_{\rm IC} \approx 5
\times 10^{-20} I_s T_{\rm keV} \left( \frac{\xi_{\rm CR}}{0.1} \right)
\left( \frac{n_e}{10^{-3} \cmden} \right)^2 \left( \frac{ k_B
T_{CMB}}{\epsilon_\gamma} \right)^{s/2} \cmdensec,
\label{eq:icem}
\eq
where $I_s = 1$ ($10^3$) for an $s=2$ ($s=3$) proton spectrum.  This is
shown in Figures \ref{fig:dndkicn2} and \ref{fig:dndkicn3} by the
dotted lines.  Clearly, the IC emission \emph{from the positrons
themselves} is a substantial contaminant, and in fact hides the
annihilation radiation completely for relatively shallow proton
spectra.  Thus, only if the protons are accelerated by very weak
shocks (so that $s \sim 3$) is the annihilation radiation from the
secondary positrons observable even in principle.

In the steady injection case, $(dn_\gamma/d\epsilon_\gamma d t) \propto
N_{p,\,0} n_e \propto \xi_{\rm CR} n_e^2$.  The ambient density also
determines the energy at which the 
continuum begins to decline rapidly; the break is located at
$\epsilon_\gamma \approx \gamma_{\rm pk} m_e c^2$, where $\gamma_{\rm pk}$
is the location of the peak of $N_+(\gamma)$ and is determined by the
efficiency of Coulomb cooling.  The temperature affects the normalization
of the continuum and line ($\dot{n}_{\rm line} \propto T_e$) as well as
the determination of the width of the line component, as described in the
previous subsection.

\section{ Discussion}
\label{conc}

We have calculated the signals expected from positron annihilation in
galaxy clusters for positrons injected as primaries by embedded radio jets
and those produced as secondaries in collisions between cosmic rays and
thermal protons in the cluster.  The former case is of interest in
constraining the matter content of relativistic jets, while the latter is
of interest in measuring the cosmic ray content of the ICM.  The positron
annihilation line is particularly interesting for these purposes because
typical cluster temperatures $k_B T_e \sim 1$--$10 \keV$ are just in the
range in which positrons annihilate efficiently without forming
positronium.  We calculate the annihilation rates assuming that the cooling
positrons mix efficiently with the ambient medium on cosmological
timescales.

\begin{figure}[t]
\plotone{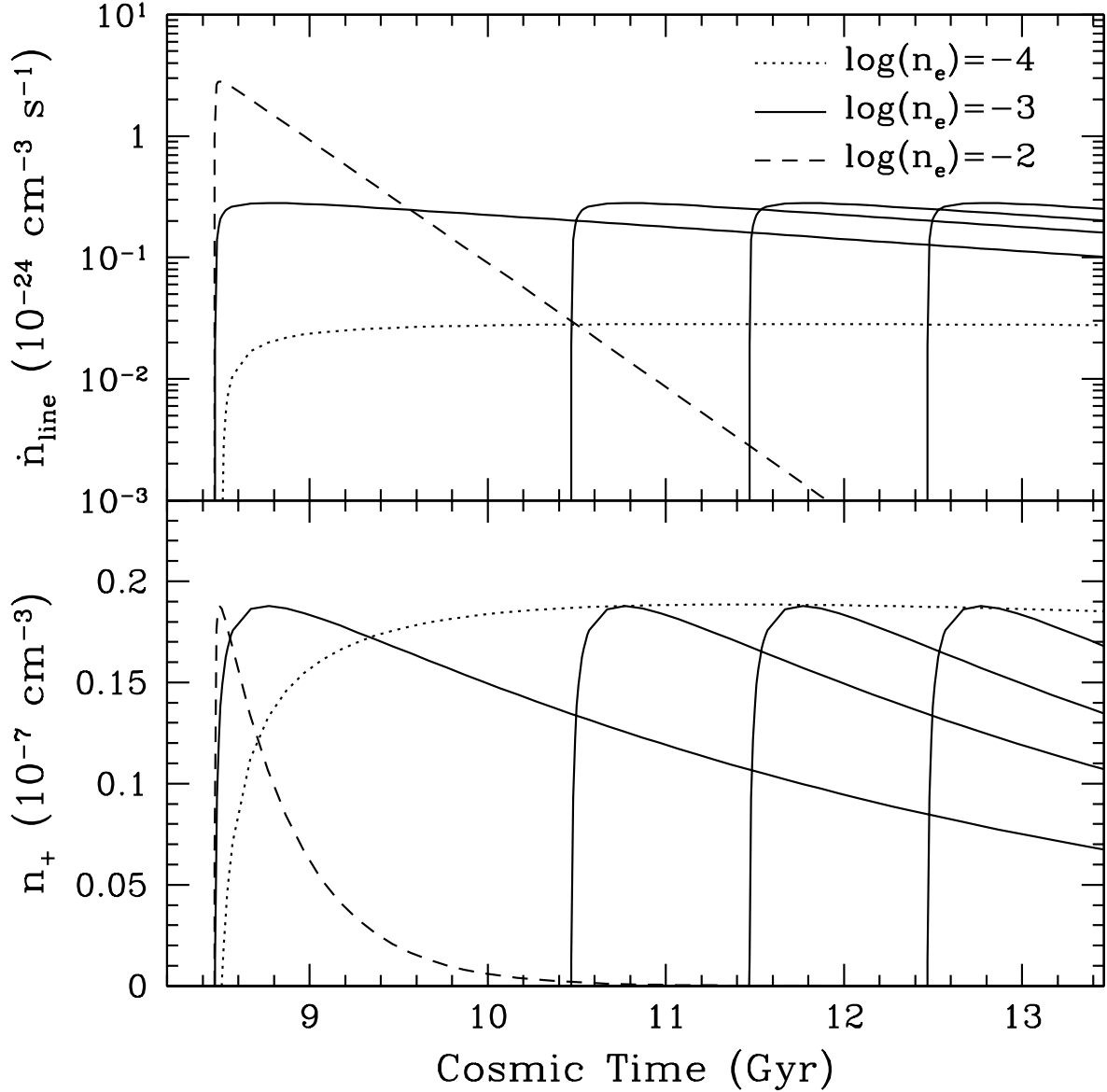}
\caption{ The evolution of the thermalized positron density $n_+$
(\emph{bottom panel}) and of the emissivity in the annihilation
line $\dot{n}_{\rm line}$ (\emph{top panel}) as a function of cosmic
time for an $s=2$ positron injection spectrum in the AGN case. 
Shown are results for $n_e = 10^{-3} \cmden$ with $t_0 - t_i =
1,\,2,\,3,$ and $5 \times 10^9 \yr$ from left to right (\emph{solid
curves}),  $n_e = 10^{-4} \cmden$ with $t_0 - t_i = 5 \times 10^9 \yr$
(\emph{dotted curve}), and $n_e = 10^{-2} \cmden$ with $t_0 - t_i = 5
\times 10^9 \yr$ (\emph{dashed curve}). All curves assume standard
cluster parameters (with the exception of $n_e$; see text). } 
\label{fig:linetimes2}
\end{figure}

\begin{figure}[t]
\plotone{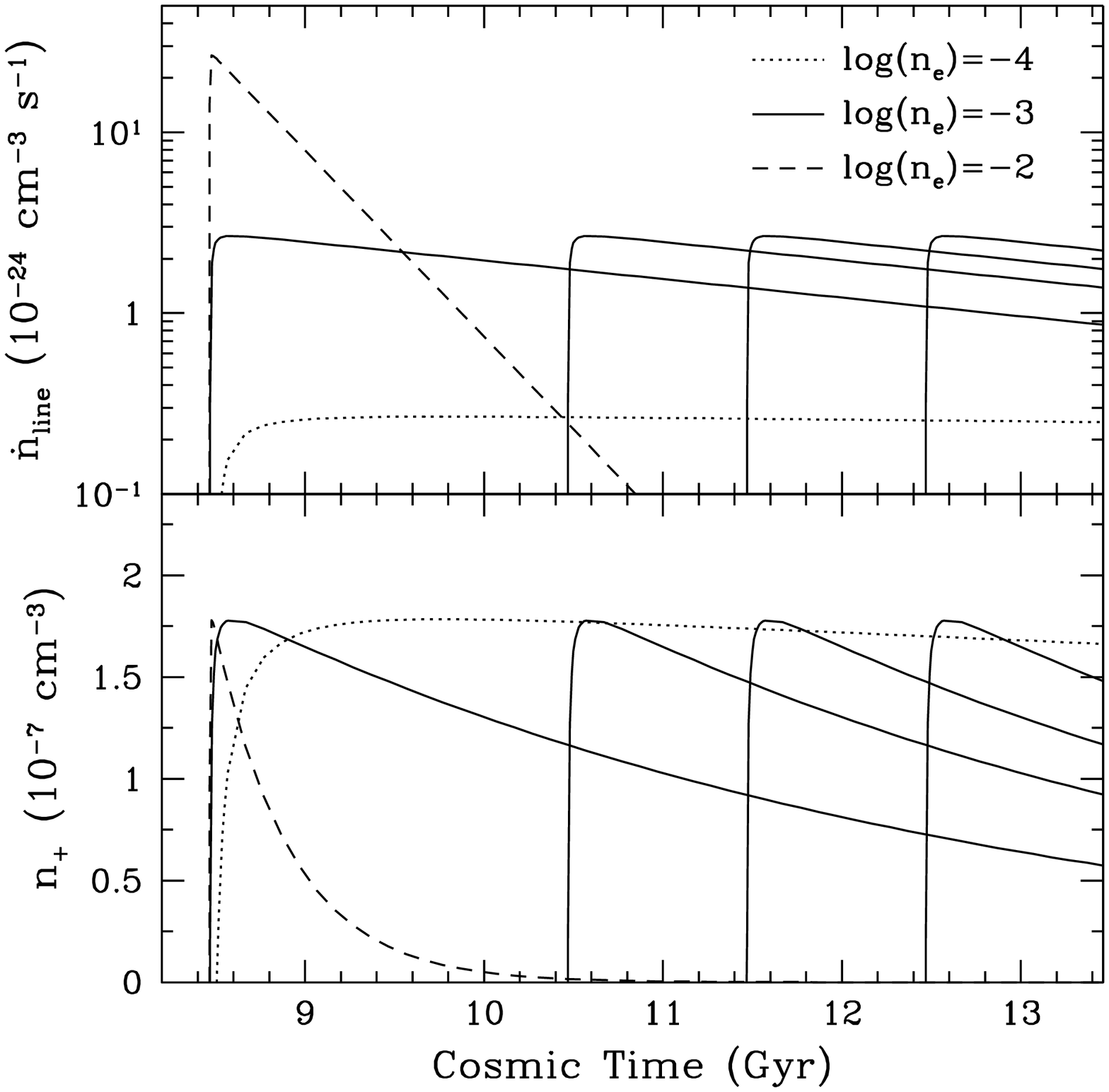}
\caption{ Same as Figure \ref{fig:linetimes2}, but with an $s=3$
positron injection spectrum.}
\label{fig:linetimes3}
\end{figure}

For positron injection by a single AGN, the peak emissivity in the
annihilation line is 
\begin{eqnarray}
\dot{n}_{\rm line,\, single} & \approx & 2 \times 10^{-25} X_s^{\rm
AGN} \left( \frac{n_e}{10^{-3} 
\cmden} \right) \left( \frac{\xi_{\rm AGN}}{0.1} \right) 
\left( \frac{r_{\rm mix}}{200 \kpc} \right)^{-3} \nonumber \\ 
\, & \, & \times \left( \frac{L_K}{10^{45} \ergs} \right) \left(
\frac{\tau}{10^8 \yr} \right) \cmdensec,
\qquad \mbox{\emph{(single AGN event)}}
\label{eq:agnlineem}
\end{eqnarray}
with $X_s^{\rm AGN} = 1$ $(10)$ for an $s=2$ $(3)$ positron injection
spectrum.  The dependence on $s$ appears primarily because, with the
normalization procedure described in \S \ref{agnsoln}, the total number of
$e^+$ produced increases as $s$ increases.  We note that rich clusters can
have core densities $n_e \sim 10^{-2} \cmden$ and $\xi_{\rm AGN}$ may reach
a value of several tens of percent for jets with a low bulk Lorentz factor.
The time lag between injection and peak annihilation depends principally on
the electron density $n_e$ in the cluster core: the characteristic
annihilation time of the positron population is $\tau_{\rm ann} \sim 4
\times 10^9 (n_e/10^{-3}~{\rm cm^{-3}})^{-1} \yr$, and the emissivity
maintains the peak level for roughly this time period (see Figures
\ref{fig:linetimes2} and \ref{fig:linetimes3}).  Therefore, although dense
clusters produce the strongest signals, they fade relatively quickly.

The longevity of the positron population in typical clusters indicates
that the signal can be enhanced if we consider either multiple
injection epochs or multiple AGN in a single cluster.  A model of the
former type has been recently advocated by \citet{bohringer}
as a way of balancing the cooling flow radiation in some clusters.
These authors suggest quasi-periodic mechanical energy injection from
the central galaxy throughout the lifetime of the cluster, with
injection epochs lasting $\sim 10^8 \yr$ occurring every $\Delta
t_{\rm inj} \sim 10^9 \yr$.  
In the limit in which $\tau_{\rm ann} \gg \Delta t_{\rm inj}$, the
annihilation emissivity will reach a quasi-steady state with 
$\dot{n}_{\rm line} \sim \eta^{-1} \dot{n}_{\rm line,\,single}$,
where $\eta = \Delta t_{\rm inj}/\tau_{\rm ann}$ is the period between
outbursts in units of the annihilation time of the cluster.  
Here $L_K$ and $\tau$ are to be interpreted as the mechanical luminosity
and lifetime of a single outburst.

An even more interesting possibility is positron injection by galaxies
throughout the cluster.  While the vast majority of these galaxies are not
active at the present day, recent studies have found evidence for relic
supermassive black holes in nearly all bulge-dominated galaxies
\citep{magorrian,gebhardt}.  This, together with modeling of the quasar
luminosity function, suggests that nearly all galaxies once hosted a quasar
\citep{haiman,haehnelt}.  Current observations give the relation $M_{\rm
BH} = 7.8 \times 10^7 (L_{B,\,{\rm bulge}}/10^{10} \lbsun)^{1.08} \msun$
\citep{kormendy}.  Because the observed black hole mass-bulge
luminosity relation is nearly linear, for the purposes of an estimate
it suffices to assume a linear relation and scale our results with the
total cluster core luminosity $L_{B,\,{\rm cl}}$.  A typical rich
cluster core has a total $B$-band luminosity of $L_{B,\,{\rm cl}} \sim
10^{12} \lbsun$ \citep{peebles}.  [This luminosity 
corresponds to $\sim 100 L_B^{\star}$ galaxies in the cluster, where
$L_B^{\star}$ is the characteristic luminosity of galaxies
in the Schechter function \citep{yasuda}.]
We further assume a mass-to-energy conversion efficiency
$\varepsilon_{\rm BH}$ during the black hole formation process and that a
fraction $f_K$ of this energy is released in outflows.  We expect
$\varepsilon_{\rm BH} \sim 0.1$, and observations indicate that $f_K \sim
0.1$ \citep[and references therein]{hooper,bigm}.
Therefore, if we make the extreme assumption that all injection events
occur simultaneously, the peak line emissivity would be
\begin{eqnarray}
\dot{n}_{\rm line} & \approx & 9 \times 10^{-23} X_s^{\rm AGN} \left(
\frac{n_e}{10^{-3} 
\cmden} \right) \left( \frac{\xi_{\rm AGN}}{0.1} \right) 
\left( \frac{r_{\rm mix}}{200 \kpc} \right)^{-3} \nonumber \\ 
\, & \, & \times \left( \frac{f_K}{0.1} \, \frac{\epsilon_{\rm
BH}}{0.1} \, \frac{L_{B,\,{\rm cl}}}{10^{12} \lbsun} \right)
\cmdensec, \qquad \mbox{ \emph{(multiple simultaneous AGN)}}
\label{eq:multagn}
\end{eqnarray}
with $X_s^{\rm AGN}$ defined as above.  Of course, we must keep in mind
that the quasar era peaked at $z \sim 2$ (\citealt{pei}; in our cosmology,
$t_0 - t_i \sim 10^{10} \yr$) so a substantial fraction of the population
may have annihilated before the present day.
In the opposite limit, in which the source evolution time is much
larger than the annihilation time, but in which the annihilation time
is in turn much larger than the time between injection events (or
$H_0^{-1} \gg \tau_{\rm ann} \gg \Delta t_{\rm inj}$), a quasi-steady
state will be reached as described in the previous paragraph, with
a steady annihilation emissivity of approximately $\eta^{-1}$ times
that of a typical AGN in the cluster.

\begin{figure}[t]
\plotone{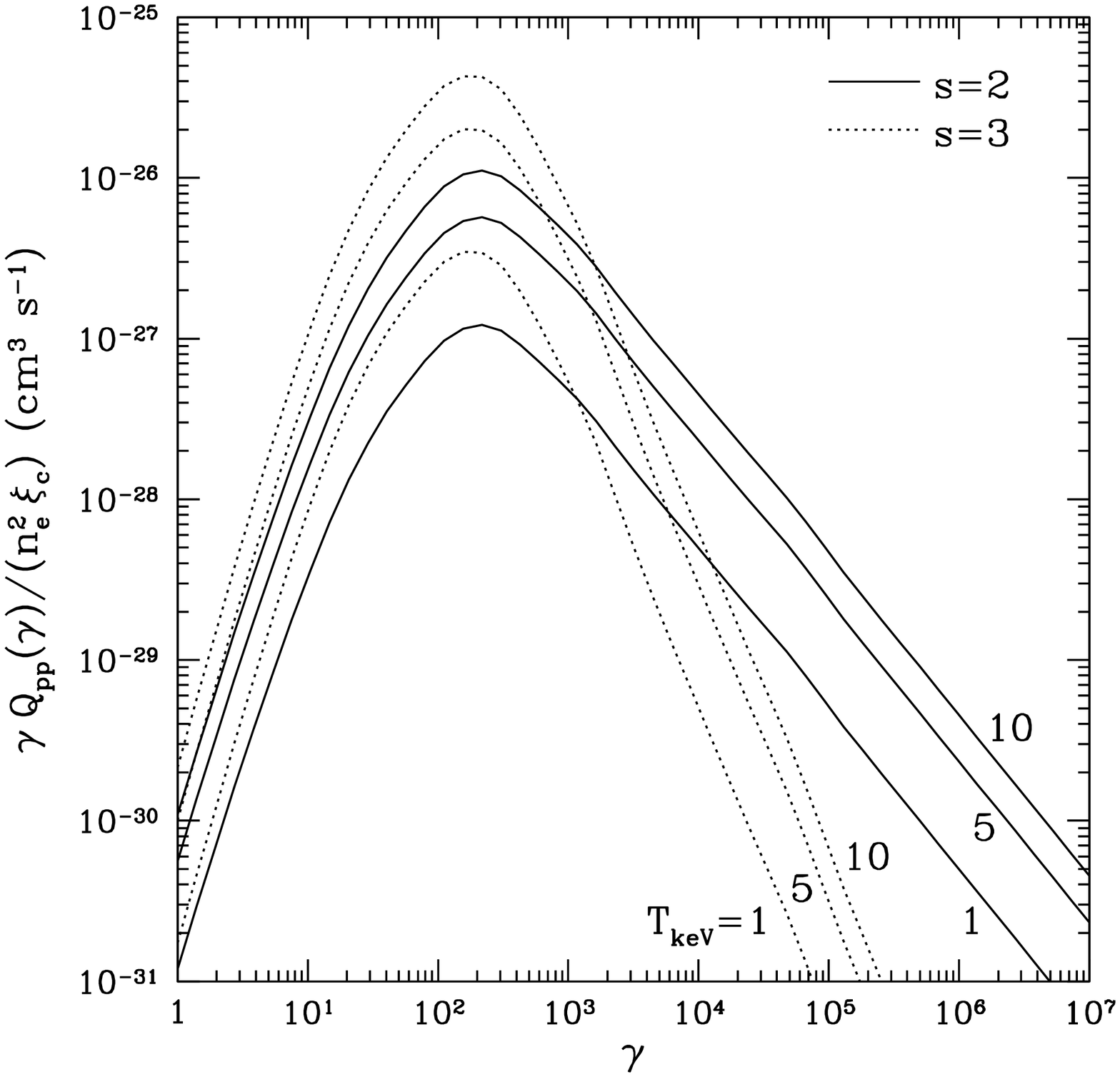}
\caption{ Normalized positron production spectrum from proton-proton
collisions, $Q_{pp}(\gamma)/(n_e^2 \xi_{\rm CR})$.
Shown are source functions for an $s=2$
(\emph{solid curves}) and an $s=3$ cosmic ray proton spectrum (\emph{dotted
curves}).  In each case the curves assume $k_B T_e = 1,\,5,$ and $10
\keV$, from bottom to top. }
\label{fig:qpp}
\end{figure}

For positron production by cosmic rays, the steady-state emissivity in the
annihilation line is 
\bq 
\dot{n}_{\rm line} \approx 10^{-30} X_s^{\rm CR}
\, T_{\rm keV} \left( \frac{n_e}{10^{-3} \cmden} \right)^2 \left(
\frac{\xi_{\rm CR}}{0.1} \right) \cmdensec, \qquad \mbox{\emph{(cosmic ray
secondaries)}}
\label{eq:crlineem}
\eq
with $X_s^{\rm CR} = 1$ $(2.5)$ for an $s=2$ $(s=3)$ proton
spectrum.  For an $s=2$ input spectrum, the energy lost to positron 
annihilation over the age of the universe is $\sim 10^{-6} U_{\rm CR}
(n_e/10^{-3} \cmden)$, where $U_{\rm CR}$ is the cosmic ray energy of
the cluster. The emission mechanism is inefficient because $\la
10^{-3}$ of the cosmic ray energy goes into the positrons and because
the characteristic initial energy of the positrons is $\gg m_e c^2$ (see
Figure \ref{fig:qpp}) so that most of the initial positron energy is lost
to cooling radiation before annihilation occurs.  We therefore see
that direct positron injection by even a weak AGN ($L_K \ga 10^{41}
\ergs$) would overwhelm the signal from secondary positrons produced
by cosmic ray protons. 

The observability of the line is of course the critical question.
The flux at earth from a cluster with a positron mixing radius $r_{\rm
mix}$ is
\bq
F_c = 8 \times 10^{-7} \left( \frac{ r_{\rm mix} }{200 \kpc}
\right)^3 \left( \frac{D}{100 \Mpc} \right)^{-2} \left(
\frac{\dot{n}_{\rm line}}{10^{-24} \cmdensec} \right) \photflux,
\label{eq:earthflux}
\eq 
where $D$ is the luminosity distance to the cluster.  In the AGN
injection case, the emissivity is $\propto r_{\rm mix}^{-3}$, and so the
flux is independent of the mixing scale (rather it depends on the total
number of positrons injected by the AGN).  For steady production by cosmic
ray protons, the emissivity is independent of the volume of the cluster
core and the flux scales in proportion to the mixing volume.

\begin{figure}[t]
\plotone{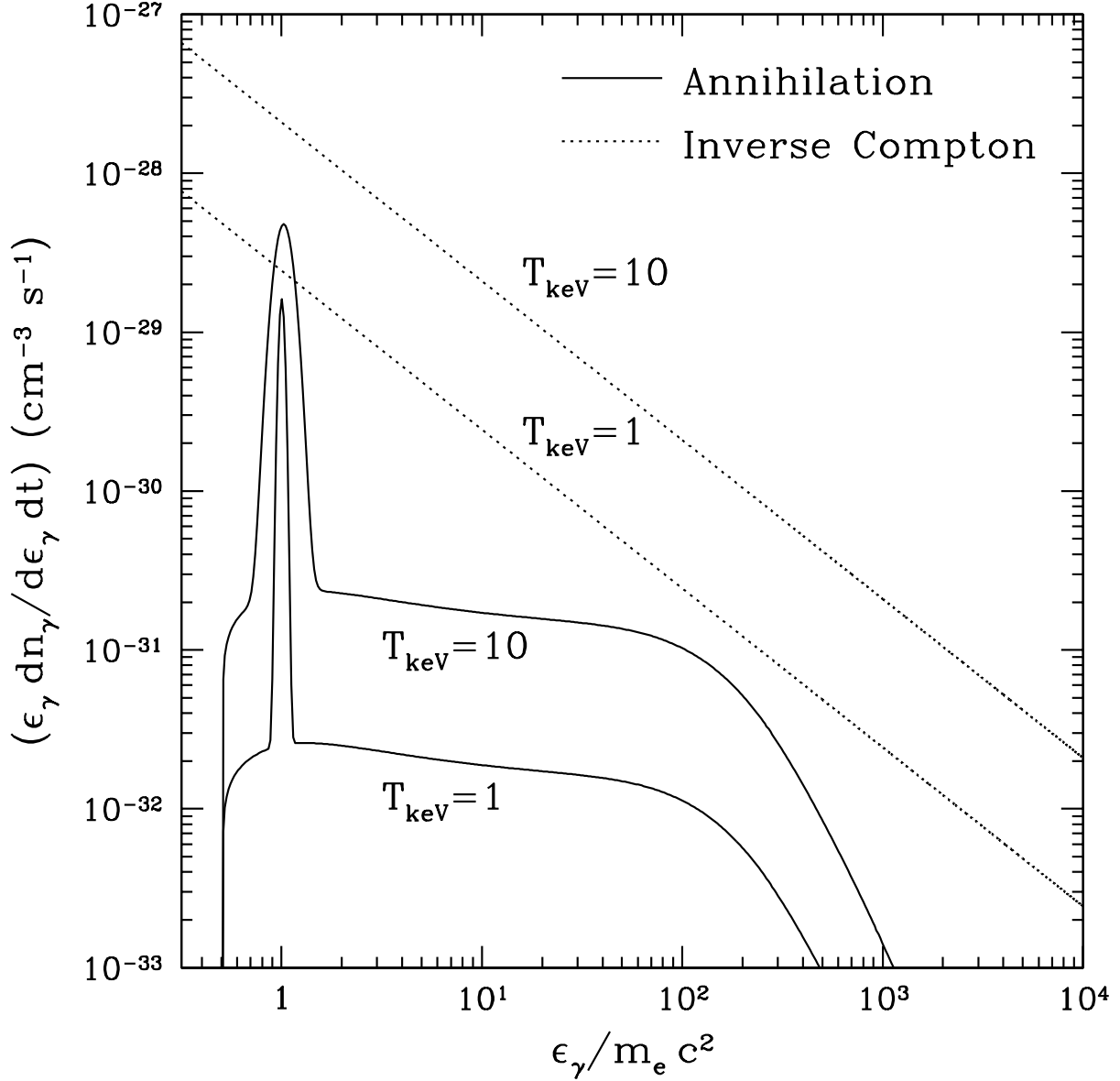}
\caption{ The spectral emissivity $(d n_\gamma/d\epsilon_\gamma\,dt)$ 
assuming steady $Q_{pp}(\gamma)$ for an $s=2$ proton spectrum.
\emph{Solid curves} show the annihilation emissivity for a clusters at
$k_B T_e = 1 \keV$ (bottom) and $k_B T_e = 10 \keV$ (top).
\emph{Dotted curves} show the inverse Compton emissivity determined
from $N_+(\gamma)$ for the same temperatures.  All curves assume $n_e =
10^{-3} \cmden$, $B = 3 \microgauss$, and $\xi_{\rm CR} = 0.1$.  }
\label{fig:dndkicn2}
\end{figure}

\begin{figure}[t]
\plotone{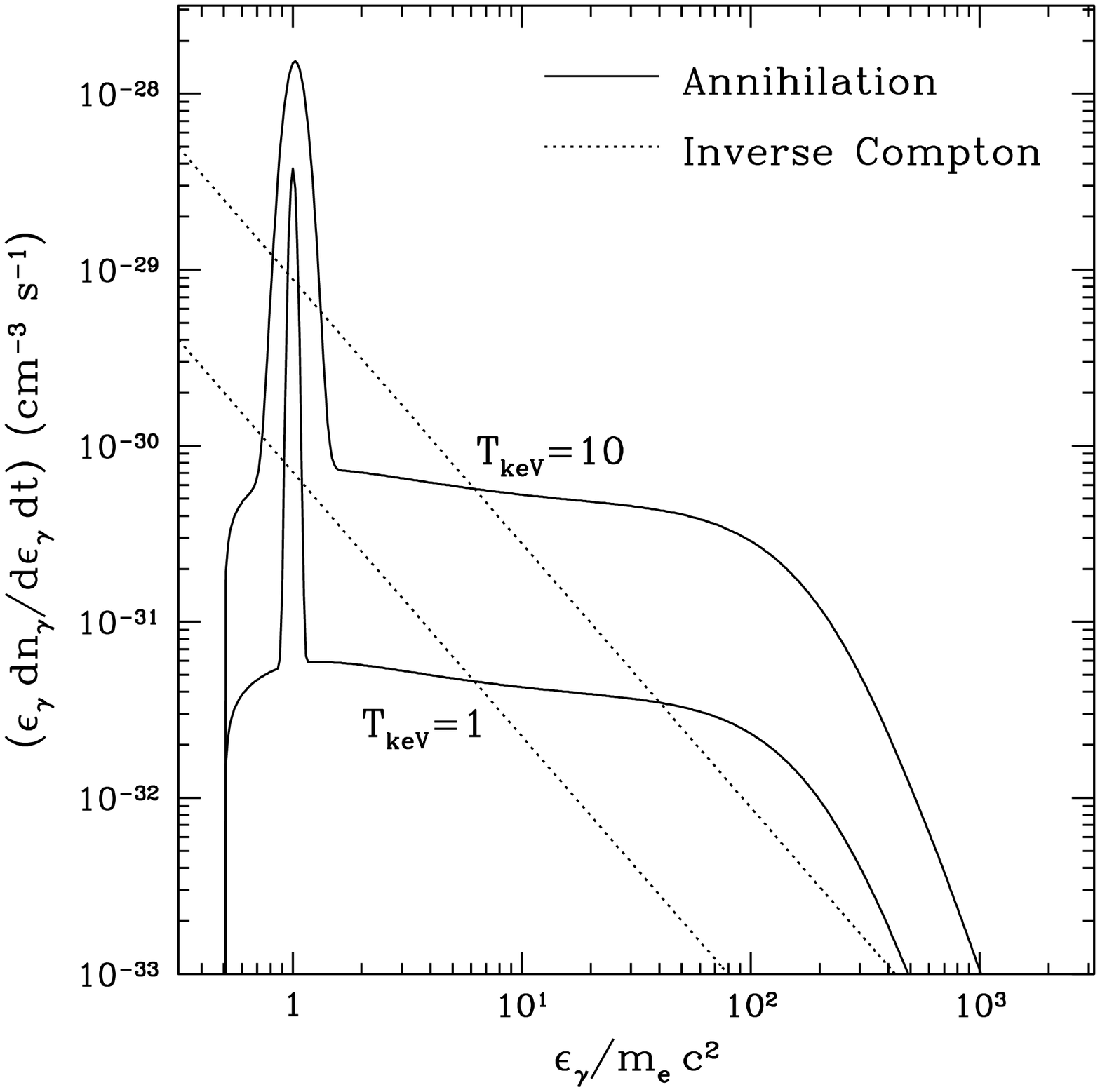}
\caption{ Same as Figure \ref{fig:dndkicn2}, but with an $s=3$
proton injection spectrum. }
\label{fig:dndkicn3}
\end{figure}

Before discussing the prospects for detection of this signal with
upcoming instruments, we must consider possible contaminating 
backgrounds.  First, as described in the previous subsection, the
relativistic positrons (and electrons) with 
$\gamma \ga 10^4$ in the cluster generate IC radiation at $511 \keV$.
However, cooling depletes this population rapidly;
after a time interval $t - t_i \sim 1.5 \times 10^8 \yr$, the maximum 
Lorentz factor $\gamma_{\rm max} < 10^4$ and the electrons and
positrons can no longer produce IC radiation at 511 keV.  However,
in the case of steady injection from cosmic ray protons, the IC background
originates from the equilibrium distribution itself.
Figures \ref{fig:dndkicn2} and \ref{fig:dndkicn3} show that in this case
the IC background can overwhelm the annihilation line unless the
protons have a sufficiently steep ($s \sim 3$) injection spectrum.

Second, we must also consider the diffuse extragalactic background, for
which the flux at $\epsilon_\gamma \approx 511 \keV$ is $F_{\rm bkgd} \sim
2 \times 10^{-5} \bkgdflux$ \citep{watanabe}.  For a cluster of a fixed
size, the ratio of the line flux from the cluster to the background
flux is independent of cluster distance so long as $z\ll 1$,
\bq 
\frac{F_c}{F_{\rm bkgd}} \sim 10^2 \, T_{\rm keV}^{-1/2}
\left( \frac{ r_{\rm mix}}{200 \kpc} \right) 
\left( \frac{\dot{n}_{\rm line}}{10^{-24} \cmdensec}
\right).
\label{eq:fluxratio}
\eq
Thus, the expected signal from AGN-injected positrons is well above
the diffuse background. However, the signal from secondary positrons
generated by cosmic rays will be hidden by the background.

While the prospects for observing the annihilation of secondary
positrons in the foreseeable future are small (even if the
annihilation line can be observed over the IC background), positrons
injected by AGN may soon be detectable with spaceborne instruments.
The INTEGRAL satellite\footnote{ See
http://astro.estec.esa.nl/SA-general/Projects/Integral/integral.html}, 
expected to be launched in October 2002, will have spectral
capabilities in the energy range of interest.  The SPI instrument is
expected to have a $3 \sigma$ line sensitivity $\sim 5.1 \times
10^{-6} \photflux$ given an integration time of $10^6 \sec$, but its
poor angular resolution ($2.5\arcdeg$) may lead to background
contamination.  The IBIS instrument, with $12\arcmin$ resolution, is
better suited to cluster detection, but it has a $3\sigma$ line
sensitivity of only $\sim 2 \times 10^{-5} \photflux$ (again for an
integration time of $10^6 \sec$).  An even more powerful search could be
conducted with  EXIST\footnote{ See
http://exist.gsfc.nasa.gov }, a proposed all-sky
hard X-ray survey mission.  It has an expected $5 \sigma$ line 
sensitivity of $\sim 5 \times 10^{-6}\photflux$ in the relevant energy
range (assuming an integration time of $10^7 \sec$, the mean exposure
time planned for any point on the sky in the mission), and it has an
excellent angular resolution of $5\arcmin$.  If the positrons are
injected by AGN, these sensitivity limits are close to the signal we
predict for nearby ($D \la 100 \Mpc$) clusters with powerful AGN, multiple
injection epochs/galaxies, or steep injection spectra.  Deep exposures
with INTEGRAL or statistical analyses taking advantage of the full sky
coverage of EXIST may reveal weaker sources as well.

A particularly interesting source is the nearby Virgo cluster, at a
distance $\sim 20 \Mpc$.
In the AGN scenario, the annihilation signal in
Virgo is comparable to the detection threshold of all three
instruments listed above.  The expected peak flux is 
\begin{eqnarray}
F_{\rm Virgo} & \approx & 1.3 \times 10^{-6} X_s^{\rm AGN} \eta^{-1}
\left( \frac{n_e}{3 \times 10^{-3} \cmden} \right) 
\left( \frac{\xi_{\rm AGN}}{0.1} \right)
\nonumber \\
\, & \, & \times \left( \frac{L_K}{10^{44} \ergs} \right) 
\left(\frac{\tau}{10^8 \yr}
\right) \photflux,
\label{eq:m87flux}
\end{eqnarray}
where we have used fiducial values for the luminosity of M87 estimated by
\citet{bohringer} and for $n_e$ by \citet{nulsen}.  Here $\eta^{-1}$
represents the contribution from past AGN phases of cluster galaxies [see
the discussion accompanying equation (\ref{eq:multagn})]; in the best case,
it could represent an enhancement of more than an order of magnitude.  If
$r_{\rm mix}$ is large, the high resolution instruments may even be able to
map spatial variations in the positron component (and indicate whether the
positrons are injected solely by the central galaxy or by a larger number
of galaxies during the quasar era).

Continuum $\gamma$-ray emission from M87, the dominant galaxy in Virgo, may
contaminate the annihilation line signal.  To date there have been no
observations of Virgo in the relevant energy range.  To estimate the
contamination from M87, we assume that observed power-law spectrum in the
$2$--$10 \keV$ range extends to $511 \keV$; using the recent observations
of \citet{bohr-m87}, we expect M87 to produce a flux in the spectral regime
of the annihilation line of $\sim 4 \times 10^{-8} T_{\rm keV}^{1/2}
\photflux$ (the temperature of the core of Virgo varies with radius between
$k_B T_e = 1$--$3 \keV$; see \citealt{bohr-m87}).  Thus, even if the
continuum emission from M87 cannot be removed through the use of a
high-angular resolution instrument, the annihilation line should still
be visible.  Of course, because M87 is still active, any positrons
produced in the current outburst phase have most likely not had
sufficient time to cool.  We would therefore expect annihilation line
emission only if there is a relic population of positrons either from
earlier outbursts of M87 or from the other galaxies in Virgo.

Another interesting source is Centaurus A, the nearest ($D \approx 3.5
\Mpc$) bright radio galaxy to the Milky Way.  The radio structure is
one of the largest in both apparent and absolute size, covering
an area $8\arcdeg \times 4\arcdeg$ on the sky (or $480 \kpc \times 240
\kpc$) and is therefore a very attractive candidate for observations
\citep{israel}.  The expected flux in the positron annihilation line
is 
\begin{eqnarray}
F_{\rm CenA} & \approx & 10^{-5} X_s^{\rm AGN} \eta^{-1}
\left( \frac{n_e}{10^{-2} \cmden} \right) 
\left( \frac{\xi_{\rm AGN}}{0.1} \right)
\nonumber \\
\, & \, & \times \left( \frac{L_K}{10^{43} \ergs} \right) 
\left(\frac{\tau}{1.4 \times 10^8 \yr}
\right) \photflux,
\label{eq:cenflux}
\end{eqnarray}
where we have scaled $L_K$ to the approximate bolometric luminosity of the
Centaurus A nuclear source \citep{chiaberge} and $\tau$ to the minimum
source age estimated by \citet{saxton}.  Recent data from the Chandra X-ray
Observatory indicates that the gas density around Centaurus A is $n_e
\approx 10^{-2} (r/5 \kpc)^{-1.33}~{\rm cm^{-3}}$ at $0.5 \kpc \ll r \la 10
\kpc$ \citep{kraft}.  The signal is potentially stronger than that of M87;
however, several caveats are in order.  First, the gas near the central
source has a temperature $k_B T_e \sim 0.275\pm 0.03 \keV$ \citep{kraft},
within the regime in which positronium formation begins to play a
significant role.  More importantly, the scale of the radio emission
suggests that a large fraction of the positrons escape the interstellar
medium of Centaurus A and mix with the intragroup medium at $r \ga 100
\kpc$.  The inferred virial temperature of the group is only $\sim 0.07
\keV$ (based on the observed velocity dispersion of group galaxies;
\citealt{vandenbergh}).  At this temperature, the positronium formation
rate is approximately equal to the free annihilation rate
\citep{crannell}. The gas density will also be much lower at large radii
(if the observed power-low decline at $r \sim 10 \kpc$ continues, then $n_e
\sim 2 \times 10^{-4} \cmden$ at $r \sim 100 \kpc$).  Therefore it is
unclear whether annihilation in this region can produce a strong line.  In
addition, because the source is still active, the positrons may not yet
have had time to cool or mix sufficiently with the surrounding gas.
Finally, because the Centaurus A group contains only $\sim 30$ galaxies
\citep{vandenbergh}, additional positron enrichment from other AGNs is
likely to be small.

In summary, we have shown that, although there are a variety of mechanisms
for producing positrons in galaxy clusters, only direct injection by AGNs
is likely to produce an observable signal.  Therefore, we argue that a
positive detection of positron annihilation lines from clusters suspected
of harboring dormant AGNs would be a robust indication that radio jets
contain an $e^+e^-$ pair plasma.

\acknowledgements

We thank J. Grindlay for helpful comments on the manuscript and
R. Kraft for providing unpublished data on Centaurus A.  This work
was supported in part by NASA grants NAG 5-7039, 5-7768, and NSF grants
AST-9900877, AST-0071019 for AL.  SRF acknowledges the support of an NSF
graduate fellowship.

\end{document}